\documentclass{jfm}

\usepackage{graphicx}
\usepackage{newtxtext}
\usepackage{newtxmath}
\usepackage{natbib}
\usepackage{hyperref}
\hypersetup{
    colorlinks = true,
    urlcolor   = blue,
    citecolor  = black,
}

\newcommand{\RomanNumeralCaps}[1]
\usepackage{algpseudocode}
\usepackage{algorithm}

\newcommand{\bmit}[1]{\bm{\mathit{#1}}}

\usepackage{epstopdf, epsfig}
\usepackage{bm}
\usepackage{upgreek}
\usepackage{xcolor}

\shorttitle{An empirical model of noise sources in subsonic jets}
\shortauthor{U. Karban et al.}

\title{An empirical model of noise sources in subsonic jets}

\author{U. Karban\aff{1}
  \corresp{\email{ukarban@metu.edu.tr}}, B. Bugeat\aff{2}, A. Towne\aff{3}, L. Lesshafft \aff{4}, A. Agarwal \aff{2}, P. Jordan \aff{5}}

\affiliation{\aff{1} {Department of Aerospace Engineering, Middle East Technical University, Ankara, Turkey}
\aff{2} Department of Engineering, Trumpington St, Cambridge CB2 1PZ, U.K.
\aff{3}{Department of Mechanical Engineering, University of Michigan, Ann Arbor, MI 48109, USA}

\aff{4}{Laboratoire d’Hydrodynamique, CNRS / \'{E}cole polytechnique, Institut Polytechnique de Paris, Palaiseau, France}

\aff{5}D\'{e}partement Fluides, Thermique, Combustion, Institut PPrime, CNRS-University of Poitiers-ENSMA, France
}

\begin{document}

\maketitle

\begin{abstract}
Modelling the noise emitted by turbulent jets is made difficult by their acoustic inefficiency: only a tiny fraction of the near-field turbulent kinetic energy is propagated to the far field as acoustic waves.  As a result, jet-noise models must accurately capture this small, acoustically efficient component hidden among comparatively inefficient fluctuations.  In this paper, we identify this acoustically efficient near-field source from large-eddy-simulation data and use it to inform a predictive model.
Our approach uses the resolvent framework, in which the source takes the form of nonlinear fluctuation terms that act as a forcing on the linearized Navier-Stokes equations.  First, we identify the forcing that, when acted on by the resolvent operator, produces the leading spectral proper orthogonal decomposition modes in the acoustic field for a Mach 0.4 jet.  Second, the radiating components of this forcing are isolated by retaining only portions with a supersonic phase speed.  This component makes up less than 0.05\% of the total forcing energy but generates most of the acoustic response, especially at peak (downstream) radiation angles.  Finally, we propose an empirical model for the identified acoustically efficient forcing components.  The model is tested at other Mach numbers and flight-stream conditions and predicts noise within 2 dB accuracy for a range of frequencies, downstream angles, and flight conditions. 
\end{abstract}

\section{Introduction} \label{sec:intro}

{Jet noise is one of the most studied problems in aeroacoustics, thanks largely to Lighthill's  theoretical framework that allows a connection to be made between the stochastic, nonlinear, vortical motions of a turbulent jet and the irrotational, linear fluctuations of the resulting acoustic field \citep{lighthill_prs_1952}. Such a framework is of interest given that there is no rigorous means by which to decompose a turbulent field into acoustic and non-acoustic components, and, therefore, no rigorous means by which to uniquely define the source of sound. }
{Lighthill's approach---an exact rearrangement of the nonlinear Navier-Stokes system into an inhomogeneous equation comprising a linear wave operator driven by a nonlinear source term---was the first of many such reorganisations of the Navier-Stokes equations \citep{curle_rspa_1955,fwilliams_rsta_1963,powell_jasa_1964,phillips_jfm_1960,
lilley1974noise,howe_jfm_1975,doak_ap_1995,goldstein_2003}.} 
All these formulations distinguish acoustic propagation from acoustic generation in a specific way, and they capture typical propagation effects like refraction in non-uniform flow with varying success.

{In parallel to these developments in aeroacoustics, similar concepts were being investigated for the study of turbulence. \cite{landahl1967wave}, for instance, proposed such a framework for the description of wall pressure fluctuations beneath a wall-bounded turbulent shear flow. More recently, the same underlying idea has been leveraged for the study of coherent structures in incompressible turbulent channel flow \citep{mckeon_jfm_2010,hwang_jfm_2010}. The novelty of these recent studies derives from: (1) a  discretisation of the inhomogeneous system; (2) a casting of the problem in frequency space; and, (3) a leveraging of the tools of linear algebra to explore the link, via the linear resolvent operator, between nonlinear interactions and the state dynamics they drive. With these three steps, the inhomogeneous system is cast in a matrix input-output form, and the relationship between `forcing' and `response'---in the context of aeroacoustics, `source' and `sound'---can be explored by considering the properties of the matrix transfer function by which they are connected: singular-value decomposition of the resolvent matrix operator can reveal the physical mechanisms by which the nonlinear forcing drives the response or by which the nonlinear source drives the acoustic field.}

{This framework has substantially enhanced our understanding of coherent structures in turbulent shear flow. Indeed, it provides a long-sought theoretical grounding for their definition \citep{towne_jfm_2018}. It has been used to study coherent structures in jets  \citep{garnaud_jfm_2013,schmidt_jfm_2018,lesshafft_prf_2019,nogueira2019large,
pickering_jfm_2021} and many other flows \citep{mckeon_jfm_2010,beneddine_jfm_2016,yeh_jfm_2019,
nogueira_jfm_2021,morra_jfm_2021}. In the context of jets, these coherent structures are most-often referred to as wavepackets \citep{crighton_pas_1975,jordan_arm_2013,cavalieri_amr_2019}.}

{The study of coherent structures has also been aided by data-processing and decomposition techniques, in particular by proper orthogonal decomposition (POD) in its numerous forms \citep{lumley_jfm_1970,picard_ijhff_2000,
boree_eif_2003,jung2004downstream,tinney2008low,towne_jfm_2018}. A recent study by \cite{karban_jfm_2022} shows how resolvent analysis and extended spectral POD \citep{boree_eif_2003} may be combined to probe turbulent shear flows in new and interesting ways. The study we report here extends this work to the jet-noise problem.}

{The extension is based on the fact that the resolvent framework can be tailored to choose what is considered as input (forcing) and output (response). The choice may involve a localisation in space and/or restriction to  a limited number of dependent variables; for example, one may inquire as to the nature of the nonlinear interactions in a turbulent boundary layer that drive shear-stress fluctuations at the wall \citep{karban_jfm_2022}. The resolvent methodology may be similarly adapted to the jet-noise problem by restricting the forcing term to the region of vortical motion, and the response to the irrotational acoustic field \citep{jeun_pof_2016}.  The resolvent framework thus resembles an acoustic analogy (see the discussion in \cite{karban_jfm_2020}); and singular value decomposition of the acoustically tailored resolvent operator can give insight into the mechanisms by which the nonlinear flow interactions drive acoustic waves.}

{However, even with this rearrangement of the problem and a study of the properties of the matrix transfer function, the problem of clearly identifying the acoustically important piece of the turbulent flow remains a challenge. This is due, on one hand, to the complex, high-rank structure of the nonlinear forcing term, and, on the other, to the formidable acoustic inefficiency of unbounded turbulence: the ratio of acoustic to turbulence fluctuation energy is of order $O(10^{-3})$ as will be shown later. These problems make it extremely difficult to craft a robust model for the acoustic source, and it is this that motivates the work we undertake. Our goal is to identify, using extensive flow data provided by large-eddy simulation, the piece of the flow that drives the acoustic field and to then propose a model for this piece of the flow that is capable of capturing effects of operating condition (jet Mach number) and forward flight.}

{There are many attempts in the literature to model the source of jet noise. A number of them are based on Goldstein's generalised acoustic analogy \citep{goldstein_2003}, where a Green's function for the linearised Euler equations (LEE) is obtained analytically. The source terms in this configuration are defined as spatio-temporal correlations of nonlinear terms in the perturbation equations. Some source models in this framework are presented for various jet configurations in \cite{goldstein_jfm_2008}, \cite{karabasov_aiaa_2010}, \cite{leib_aiaa_2011}, \cite{gryazev_aiaa_2022} among others.} 

{Another group of studies to predict jet noise are called stochastic methods, where some synthetic velocity fluctuations which satisfy two-point statistics of the flow are specified, and the acoustic field is then computed either using acoustic analogies or LEE. One branch of such stochastic approaches are called `stochastic noise generation and radiation' methods, first applied to the jet noise problem by \citet{bechara_aiaa_1994}, and then modified by \citet{billson_aiaa_2004} and \citet{lafitte_aiaa_2011}. In this method, a synthetic velocity field is obtained via summation of randomly distributed spatial Fourier modes. The energy content is defined using the von Karman-Pao energy spectrum. The resulting velocity field is convected using the mean flow, and finally, the acoustic field is predicted by applying Lighthill's analogy on this field. Another stochastic approach used for jet-noise prediction is called the `random particle mesh' (RPM) method \citep{siefert_jasa_2008}, where a synthetic stream function is generated by applying solenoidal filters on random signals, which then is used to compute the source terms forcing LEE to obtain pressure fluctuations.} 

{All of the approaches discussed so far model the source correlations in the time domain. A frequency-domain model based on resolvent analysis was introduced by \cite{towne_aiaa_2017}. They provided a model function tuned using the two-point forcing correlations in the frequency domain obtained from LES data. An empirical relation using turbulent kinetic energy and dissipation rate is then provided to replace the tuning based on LES data. In another study, \citet{pickering_jasa_2021} predicted using acoustic-field data from LES how the forcing projects onto the resolvent forcing modes. They adopted the resolvent-based estimation method given in \cite{towne_jfm_2020}, but instead of the forcing statistics, they predicted the projection coefficients and then provided an empirical model for them. This way, they leveraged the linear mechanisms embedded in the resolvent operator associated with noise generation and model only the remaining nonlinearities coming from the forcing.} 
{Other studies have used a wavepacket source model for the near-field velocity correlations paired with Lighthill’s analogy to predict jet noise \citep{huerre_aiaa_1983,cavalieri_jsv_2011,cavalieri_jfm_2012,cavalieri_jfm_2014,
maia_rspa_2019,daSilva_jasa_2019}.} 

{{In this study, we follow a strategy similar to \cite{towne_aiaa_2017} to propose a data-driven source model that is used to predict noise generation in subsonic jets, but instead of modelling the two-point forcing correlations directly, we first isolate the acoustically efficient structures.} We use the resolvent-based extended spectral proper orthogonal decomposition (RESPOD) proposed by \cite{towne_aiaa_2015} and further developed by \cite{karban_jfm_2022} to perform a preliminary filtering of the resolvent forcing data. This filtering extracts the forcing subspace correlated with the axisymmetric acoustic field radiated to low polar angles. The subspace so obtained contains silent-but-correlated and sound-producing components, and a second filtering is necessary to extract the latter. The one-to-one correspondence between the SPOD modes of the acoustic field and this forcing subspace is used to show that the dominant noise generating forcing is the acoustically matched part \citep{fwilliams_rsta_1963,crighton_pas_1975,freund_jfm_2001,
cabana2008identifying,sinayoko2011flow,cavalieri_amr_2019} of the subspace. An empirical model for the sound-generating part of the field is then proposed on the basis of the acoustically matched piece of the forcing, and adapted to capture the effects of operating condition and forward flight. } {The advantage of the proposed strategy is that it leverages the versatility of the resolvent framework similar to \cite{pickering_jasa_2021} to systematically identify the dominant noise-generation mechanisms. Different from their study, here, we start by analysing the forcing data to identify the noise-generating part and then drive the resolvent operator with this refined forcing to predict the acoustic field. Given the inefficiency of turbulence to generate noise, such an identification significantly contributes to the robustness of the final empirical model, as it retains only the essential information from the forcing data.}

{The paper is organised as follows: the mathematical framework for resolvent analysis and RESPOD is revisited in \S\ref{sec:math}. The details about the numerical database and the resolvent-analysis tool are given in \S\ref{sec:numdata}. The process to identify forcing components that generate downstream jet noise is explained in \S\ref{sec:ident}. Based on these acoustically efficient forcing components, an empirical forcing model is presented in \S\ref{sec:model}, which is then adapted to include Mach-number and flight-stream effects. Concluding remarks are provided in \S\ref{sec:conc}.}

\section{Modelling framework} \label{sec:math}
{The resolvent framework is obtained by linearising the Navier-Stokes (N-S) equations and arranging them in input-output, or forcing-response, form in the frequency domain, where the input is nonlinear fluctuations and the output is the state. In the present case, we limit the response to be the acoustic pressure and aim to extract the forcing associated with this target response. We achieve this using resolvent-based extended spectral proper orthogonal decomposition (RESPOD) \citep{karban_jfm_2022}. In this section, we briefly revisit the resolvent framework and the RESPOD approach.} 

\subsection{Governing equations in resolvent form} \label{subsec:govern}
The compressible N-S equations are given in compact form as
\begin{align} \label{eq:nscompact}
\partial_t\bm{q}=\mathcal{N}(\bm{q}),
\end{align}
where $\mathcal{N}$ is the N-S operator, $\bm{q}=[\nu,\,u_x,\,u_r,\,u_\theta,\,p]^\top$ is the state vector, $\nu$ is specific volume and $p$ is pressure, $\bm{u}=[u_x,\,u_r,\,u_\theta]^\top$ is the velocity vector in cylindrical coordinates, and $x$, $r$ and $\theta$ refer, respectively, to the streamwise, radial and azimuthal directions. All variables are non-dimensionalised by the ambient speed of sound, $c_\infty$, the density, $\rho_\infty$, and the nozzle diameter, $D$. We consider a discretised system in space, for which linearisation around the mean, $\overline{\bm{q}}$, yields
\begin{align} \label{eq:forceintime}
\partial_t\bm{q}^{\prime}-\mathsfbi{A}\bm{q}^{\prime}=\bm{f},
\end{align}
where $\mathsfbi{A}=\partial_q\mathcal{N}|_{\overline{\bm{q}}}$ is the linear operator obtained from the Jacobian of $\mathcal{N}$ and $\bm{f}$ contains all remaining nonlinear terms, which are referred to henceforth as the forcing terms. Equation \eqref{eq:forceintime} is Fourier transformed and rearranged to obtain 
\begin{align} \label{eq:forceinfreq}
\hat{\bm{q}}=\mathsfbi{R}\hat{\bm{f}},
\end{align}
where the hat indicates Fourier-transformed quantities and 
\begin{align}
\mathsfbi{R}=(i\omega\mathsfbi{I}-\mathsfbi{A})^{-1}
\end{align} 
is the resolvent operator. The resolvent operator can be modified to limit the response to prescribed measurements via a linear transformation, $\mathsfbi{C}$, of the state vector,
\begin{align} 
\hat{\bm{y}}&=\mathsfbi{C}\hat{\bm{q}},
\end{align}
such that the input-output relation between forcing, $\hat{\bm{f}}$, and measurement, $\hat{\bm{y}}$, is given by,
\begin{align}
\hat{\bm{y}}&=\mathsfbi{C}\mathsfbi{R}\hat{\bm{f}}.\label{eq:modifres}
\end{align}
Throughout this paper, we will focus on the pressure in the acoustic field as our measured quantity; therefore {$\mathsfbi{C}$ is an $N_A\times5N$ matrix, that is one in the elements that correspond to pressure in the acoustic field and zero for the rest. Here, $N_A$ and $N$ denote the number of discrete points in the acoustic field and the full domain, respectively.} 

It is also possible to impose restrictions on the forcing in \eqref{eq:modifres} via a control matrix, $\mathsfbi{B}$, as
\begin{align} 
\hat{\bm{y}}_B&=\tilde{\mathsfbi{R}}\hat{\bm{f}},\label{eq:modifres2}
\end{align}
where $\tilde{\mathsfbi{R}}\triangleq\mathsfbi{C}\mathsfbi{R}\mathsfbi{B}$ denotes the modified resolvent operator. Note that, $\hat{\bm{y}}_B\neq\hat{\bm{y}}$ in general. $\mathsfbi{B}$ will later be used to identify irrelevant forcing components for the jet-noise problem: i.e. those terms which, when suppressed by $\mathsfbi{B}$, do not lead to changes in the measurement, such that $\hat{\bm{y}}_B=\hat{\bm{y}}$.

\subsection{Resolvent-based extended spectral proper orthogonal decomposition} \label{subsec:respod}
One of the goals of this study, and indeed one of the broader goals of jet-noise modelling, is to obtain simplified representations or models of the nonlinear interactions that underpin jet noise. One such approach is to search for a useful rank reduction. A known trait of turbulent jets and their sound is the marked difference in complexity between the turbulence and acoustic fields. {This difference implies that, given a discretised turbulent jet database, it may be possible to represent the acoustic field using a compact basis with a small number of vectors, yielding a low-rank system, while a substantially larger basis will be required to define the turbulent fluctuations, and thus the forcing, in the near field, yielding a high-rank system.}
A central idea underlying the approach we follow here is as follows: given the linear relation between the forcing and the target response in \eqref{eq:modifres}, the low-rank structure of the acoustic field suggests the existence of a low-rank, acoustically active forcing subspace. It is necessary to identify this subspace, because it is there that modelling work can be done. 

There exist several ways to identify the forcing associated with a given response. A detailed analysis was provided in \cite{karban_jfm_2022}, where a method referred to as `Resolvent-based Extended Spectral Proper Orthogonal Decomposition' (RESPOD) was used to achieve the abovementioned identification. RESPOD is based on the extended proper orthogonal decomposition presented by \cite{boree_eif_2003} and is related to spectral proper orthogonal decomposition (SPOD) \citep{lumley_jfm_1970,picard_ijhff_2000,towne_jfm_2018}. The aim in RESPOD is to find a forcing mode, $\bmit{\chi}^{(p)}$, that is correlated with the $p^{\textrm{th}}$ SPOD mode, $\bmit{\psi}^{(p)}$, of the measured response, $\hat{\bm{y}}$. It was first presented in \cite{towne_aiaa_2015} and later discussed in \cite{karban_jfm_2022} to identify the forcing structures that generate wall-attached eddies. Here, we briefly review the method highlighting how it can be adapted to find the low-rank forcing subspace associated with sound generation.

For a given ensemble of realizations $\hat{\mathsfbi{Y}}=[\hat{\bm{y}}_1\cdots\hat{\bm{y}}_P]$ of an $N$ dimensional discretised system, where $P$ is the number of Fourier realizations, SPOD involves eigen-decomposition of the CSD matrix $\hat{\mathsfbi{S}}\triangleq\hat{\mathsfbi{Y}}\hat{\mathsfbi{Y}}^H$, 
\begin{align} \label{eq:pqqeig}
\hat{\mathsfbi{S}} = \hat{\bmit{\mathit{\Psi}}}\hat{\bmit{\Lambda}}\hat{\bmit{\Psi}}^H,
\end{align}
where the eigenvectors, $\hat{\bmit{\Psi}}$, and eigenvalues, $\hat{\bmit{\Lambda}}$, of $\hat{\mathsfbi{S}}$ are the SPOD modes and gains, respectively. An alternative way to obtain the SPOD modes, as shown  in \cite{towne_jfm_2018}, is to perform the eigendecomposition
\begin{align} \label{eq:qhq}
\hat{\mathsfbi{Y}}^H\mathsfbi{W}\hat{\mathsfbi{Y}}=\hat{\bmit{\Theta}}\hat{\bmit{\Lambda}}\hat{\bmit{\Theta}}^H,
\end{align}
where $\mathsfbi{W}$ is a positive-definite weight matrix, and $\hat{\bmit{\Theta}}$ is a matrix containing the eigenmodes of $\hat{\mathsfbi{Y}}^H\mathsfbi{W}\hat{\mathsfbi{Y}}$.  The eigenmodes, $\hat{\bmit{\Psi}}$ and $\hat{\bmit{\Theta}}$, are related as
\begin{align} \label{eq:quickspod}
    \hat{\bmit{\Psi}}=\hat{\mathsfbi{Y}}\hat{\bmit{\Theta}}\hat{\bmit{\Lambda}}^{-1/2},
\end{align}
or alternatively as
\begin{align} \label{eq:quickspod2}
    \hat{\bmit{\Theta}}=\hat{\mathsfbi{Y}}^H\mathsfbi{W}\hat{\bmit{\Psi}}\hat{\bmit{\Lambda}}^{-1/2}.
\end{align}
Equation \eqref{eq:quickspod} indicates that it is possible to obtain the SPOD modes as a linear combination of the realizations. Writing \eqref{eq:modifres2} with $\mathsfbi{B} = \mathsfbi{I}$ for the ensemble of realizations as
\begin{align} \label{eq:resolventensemble}
    \hat{\mathsfbi{Y}}=\tilde{\mathsfbi{R}}\hat{\mathsfbi{F}},
\end{align}
where $\hat{\mathsfbi{F}}\triangleq[\hat{\bm{f}}_1\cdots\hat{\bm{f}}_P]$ is the matrix of the forcing realisations, and multiplying \eqref{eq:resolventensemble} by $\hat{\bmit{\Theta}}\hat{\bmit{\Lambda}}^{-1/2}$ yields
\begin{align} \label{eq:resolventspod}
    \hat{\bmit{\Psi}}=\tilde{\mathsfbi{R}}\hat{\mathsfbi{F}}\hat{\bmit{\Theta}}\hat{\bmit{\Lambda}}^{-1/2}.
\end{align}
Equation \eqref{eq:resolventspod} can be written for the $p^{\textrm{th}}$ SPOD mode by extracting the corresponding columns in the matrices, $\hat{\bmit{\Psi}}$, $\hat{\bmit{\Theta}}$ and $\hat{\bmit{\Lambda}}$,
\begin{align} \label{eq:resolventspodp}
    \hat{\bmit{\psi}}^{(p)}=\tilde{\mathsfbi{R}}\hat{\mathsfbi{F}}\hat{\bmit{\theta}}^{(p)}{\lambda^{(p)}}^{-1/2},
\end{align}
where $\hat{\bmit{\theta}}^{(p)}$ denotes the $p^{\textrm{th}}$ column in $\hat{\bmit{\Theta}}$ and $\lambda^{(p)}$ denotes the $p^{\textrm{th}}$ diagonal element in $\hat{\bmit{\Lambda}}$. We then define the RESPOD mode of the forcing, $\hat{\bmit{\chi}}^{(p)}$, as
\begin{align} \label{eq:optfrc1}
\hat{\bmit{\chi}}^{(p)}\triangleq\hat{\mathsfbi{F}}\hat{\bmit{\theta}}^{(p)}{\lambda^{(p)}}^{-1/2}.
\end{align}
Following \cite{boree_eif_2003}, it can be shown that the RESPOD mode, $\bmit{\chi}^{(p)}$, contains all the forcing components correlated with the SPOD mode, $\bmit{\psi}^{(p)}$. Furthermore, \eqref{eq:resolventspodp} indicates that the two modes are connected via the resolvent operator as
\begin{align}
\hat{\bmit{\psi}}^{(p)}=\tilde{\mathsfbi{R}}\hat{\bmit{\chi}}^{(p)}.
\end{align} 

The ability to identify a RESPOD mode of the forcing with each SPOD mode of the response implies, for the jet-noise problem, that one can use this approach to identify the low-rank forcing subspace that is correlated with the low-rank acoustic field, and which, furthermore, generates the low-rank acoustic field when applied to the resolvent operator. 

We will use this approach to obtain a low-rank representation of the forcing that is responsible for most of the acoustic energy radiated by a turbulent jet. 
The advantage of identifying a low-rank forcing is twofold: (\emph{i}) one needs to model only that piece in the forcing; (\emph{ii}) not dealing with the entire CSD matrix of the forcing is convenient in terms of computing the response, which would otherwise require multiplication of the resolvent operator by an $N\times N$ matrix, where $N$ is generally very large. Given the number of realisations, $P$, and the degree of freedom, $N$, which usually satisfy $P\ll N$, this approach provides a computationally inexpensive means of obtaining a low-rank system.

\section{Numerical databases and tools} \label{sec:numdata}
The numerical analysis in this study is conducted in two stages: (\emph{i}) identification of low-rank forcing by post-processing a large eddy simulation (LES) database, and (\emph{ii}) performing acoustic predictions by computing the response of the resolvent operator driven by the identified forcing modes. In the following subsections, we provide details about the LES database and the resolvent analysis, respectively.
\subsection{Large eddy simulation database} \label{subsec:les}

The numerical database used in this study to develop an empirical forcing model consists of LES of four subsonic jets, one at jet Mach number, $M_j\triangleq U_j/c_j=0.4$, with no flight effect and others at $M_j=0.9$ with or without flight effect. We use three other LES databases at $M_j=0.7$ with or without flight effect and at $M_j=0.8$ without flight effect to test the model. The LES was conducted using the unstructured flow solver `Charles' \citep{bres_aiaa_2017}. In all cases, the jets are isothermal and ideally expanded. Two of the cases at $M_j=0.9$ contain a flight stream at $M_\infty=0.15$ and 0.3, respectively. All jets are turbulent thanks to a synthetic forcing applied inside nozzle. Other parameters related to each simulation are tabulated in Table \ref{tab:les}, where $Re=\rho_j U_jD/\mu_j$ denotes the Reynolds number, $\mu$ denotes the dynamic viscosity, $D$ is nozzle diameter, $U$, $P$ and $T$ denote the mean streamwise velocity, pressure and temperature, respectively, c.v. stands for control volume, $dt=\tilde{dt}c_\infty/D$ and $t_{\textrm{sim}}=\tilde{t}_{\textrm{sim}}c_\infty/D$ are time step and total time of the simulation in acoustic units, where $\tilde{t}$ is the physical time, and $\Delta t$ is the time step in acoustic units used for data storage. The subscripts, $j$, $\infty$ and $0$ denote jet exit, free-stream and stagnation conditions, respectively. Throughout this paper, velocities are non-dimensionalized with the ambient speed of sound, $c_\infty$, lengths with nozzle diameter, $D$, pressure with $\rho_\infty c_\infty^2/2$, and time with $c_\infty/D$. Frequencies are reported in Strouhal number, $St=\tilde{f}D/U_j$, where $f$ is the dimensional frequency. 

The axisymmetric nature of jets renders possible decomposing the flow into azimuthal Fourier modes and analysing them separately. To facilitate azimuthal Fourier decomposition, for each case, the LES data is interpolated onto a cylindrical grid with mesh size $(N_x,N_r,N_\theta)=(656,138,128)$, where $N_x$, $N_r$ and $N_\theta$ are the number of grid points in streamwise, radial and azimuthal directions. The cylindrical grid extends in $x,r,\theta \in [0,30]\times[0,6]\times[0,2\pi]$. 

The linearised N-S equations in the time domain for the $m^{\textrm{th}}$ azimuthal Fourier mode are given as
\begin{align} \label{eq:forceintimedec}
\partial_t{\bm{q}^{\prime}}^{(m)}-\mathsfbi{A}^{(m)}{\bm{q}^{\prime}}^{(m)}=\bm{f}^{(m)},
\end{align}
where the superscript $(\cdot)$ denotes the azimuthal mode number. We limit our analysis to the acoustic field in the first azimuthal mode, $m=0$, only.

\begin{figure}
  \centerline{\resizebox{1\textwidth}{!}{\includegraphics{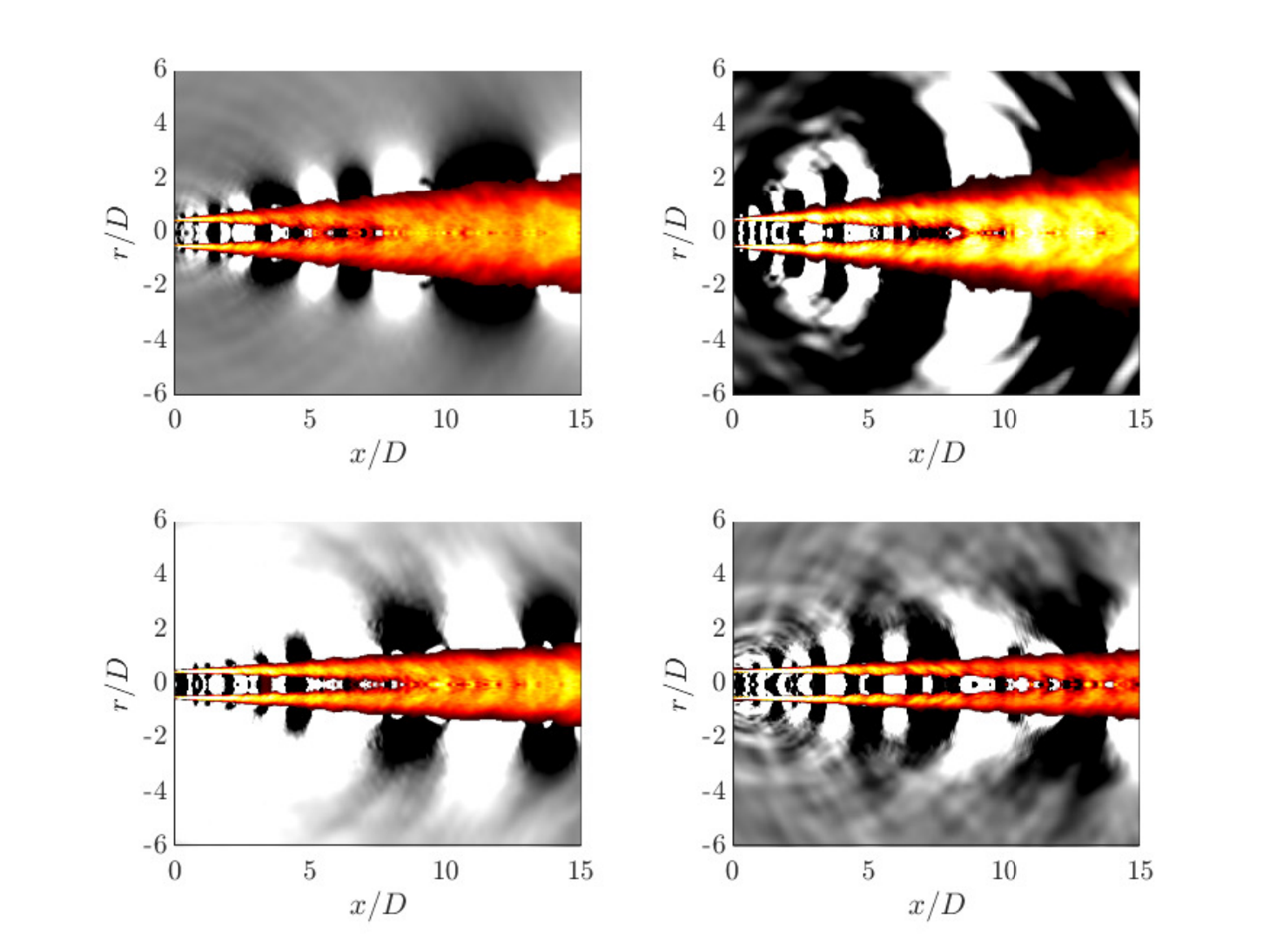}}}% Images in 100% size
\vspace{-10px}
  \caption{Snapshots of the first azimuthal Fourier mode of pressure (grey) and temperature (color) for the cases M04Mc00 (top-left), M09Mc00 (top-right), M09Mc15 (bottom-left), M09Mc30 (bottom-right). Color-scale for pressure linearly varies between $[-6\times10^{-3},6\times10^{-3}]$ for all cases. Color-scale for temperature is given as $[1,1.01]$, $[1,1.03]$, $[1,1.02]$, $[1,1.02]$ for the abovementioned four cases, respectively. }
\label{fig:snapshot}
\end{figure}

In figure \ref{fig:snapshot}, snapshots of pressure and temperature for the first azimuthal Fourier mode, $m=0$, are shown for the four cases used for tuning of the empirical model. Temperature is shown as an indicator of turbulent fluctuations in the shear layer. Pressure is saturated to show the acoustic waves propagating from the jets. It is seen that the case M09Mc00 (see table \ref{tab:les} for case abbreviations) has the strongest pressure gradients, and thus, highest noise level. Existence of the flight stream suppresses both the turbulent fluctuations and the noise generated by the jet, at a level increasing with the flight velocity. The flow fields for the remaining three cases are not shown for brevity. All seven cases are validated against experimental data. A detailed validation can be found in \cite{bres_jfm_2018} for the cases M04Mc00 and M09Mc00 and in \cite{maia_arxiv_2022} for the remaining cases. The M09Mc00 case is publicly available as part of a database for reduced-complexity modelling of fluid flows \citep{towne_arxiv_2022}.

\begin{table}
\caption{Details of the LES database. The first four and the last three cases are used to tune and test the empirical model, respectively.}
\label{tab:les}
\vspace{0.3cm}
\centering
\begin{tabular}{c c c c c c c c c c}
case & $Re$ & $M_j$ & $M_\infty$ & $P_0/P_\infty$ & $T_0/T_\infty$ & \# of c.v. & $ dt$ & $\Delta t$ & $t_{\textrm{sim}}$ \\
\hline
M04Mc00 & $0.45\times 10^6$ & 0.4 & 0.0 & 1.12 & 1.03 & $16\times 10^6$ & $1\times 10^{-3}$ & 0.1 & 3000 \\
M09Mc00 & $1.01\times 10^6$ & 0.9 & 0.0 & 1.69 & 1.16 & $16\times 10^6$ & $1\times 10^{-3}$ & 0.1  & 2000 \\
M09Mc15 & $1.01\times 10^6$ & 0.9 & 0.15 & 1.69 & 1.16 & $22\times 10^6$ & $1\times 10^{-3}$ & 0.1  & 2000 \\
M09Mc30 & $1.01\times 10^6$ & 0.9 & 0.3 & 1.69 & 1.16 & $22\times 10^6$ & $1\times 10^{-3}$ & 0.1  & 2000 \\
\hline
M07Mc00 & $0.79\times 10^6$ & 0.7 & 0.0 & 1.39 & 1.10 & $22\times 10^6$ & $1\times 10^{-3}$ & 0.1 & 2000 \\
M07Mc15 & $0.79\times 10^6$ & 0.7 & 0.0 & 1.39 & 1.10 & $22\times 10^6$ & $1\times 10^{-3}$ & 0.1  & 2000 \\
M08Mc00 & $0.90\times 10^6$ & 0.8 & 0.15 & 1.52 & 1.13 & $22\times 10^6$ & $1\times 10^{-3}$ & 0.1  & 2000 \\
\vspace{-5pt}
\end{tabular}
\end{table}

The case M04Mc00 contains both the state and the forcing data while others contain only the state data. The forcing data,  once the state data is stored, is obtained via the procedure devised in \cite{towne_phd_2016} and summarised in algorithm \ref{alg:1}.
{
\begin{algorithm}
\caption{Computing the forcing} \label{alg:1}
\begin{algorithmic}[1]
\State Calculate the state $\bm{q}$ through LES with $dt=0.001$ and store it at every 200${}^{\text{th}}$ time step.
\State Calculate and save the mean flow $\bar{\bm{q}}$.  
\State Calculate and save $\mathcal{G}(\bar{\bm{q}})$, where $\mathcal{G}$ is the nonlinear LES operator. Note that $\mathcal{G}$ is different from the N-S operator, $\mathcal{N}$, as the sub-grid scales are filtered in $\mathcal{G}$.
\State For each snapshot, calculate $\partial \bm{q}/\partial t = \mathcal{G}(\bm{q})$.
\State For each snapshot, calculate $\mathsfbi{A}\bm{q}^\prime\approx\frac{\mathcal{G}(\bar{\bm{q}} + \epsilon\bm{q}^\prime) - \mathcal{G}(\bar{\bm{q}})}{\epsilon}$, where $\epsilon$ is a sufficiently small number. 
\State Interpolate $\bm{q}$, $\partial \bm{q}/\partial t$, and $\mathsfbi{A}\bm{q}^{\prime}$ data onto the  cylindrical grid.
\State Compute the forcing in the time domain using \eqref{eq:forceintime}.
\end{algorithmic}
\end{algorithm}
}
Both the state and forcing data are Fourier transformed using blocks containing 512 snapshots in time with an overlap ratio of 75\%. To minimize spectral leakage, we use an exponential windowing function \citep{martini_arxiv_2019},
\begin{align} \label{eq:win}
W(t) = e^{n\left(4-\frac{T}{t(T-t)}\right)},
\end{align}
with $n=1$ and window size $T=512\Delta t$. The correction discussed in \cite{martini_arxiv_2019}, which is necessary to satisfy \eqref{eq:forceinfreq} when a windowing function is applied during the temporal Fourier transform (FT), is implemented while computing the forcing terms in the frequency domain. The correction is shown in \cite{nogueira_jfm_2021} to significantly improve the convergence of resolvent-based prediction of the response via \eqref{eq:forceinfreq}.

\subsection{Resolvent analysis} \label{subsec:restool}
Resolvent-based prediction of the response using the forcing data is achieved via a custom resolvent analysis code \citep{bugeat20193d}. The code uses the finite-volume method to solve the linearised N-S equations decomposed into azimuthal Fourier modes. The input-output relation in \eqref{eq:forceinfreq} is written for a given azimuthal mode, $m$, as
\begin{align} \label{eq:forceinfreqm}
\hat{\bm{q}}^{(m)}=\mathsfbi{R}^{(m)}\hat{\bm{f}}^{(m)},
\end{align}
where $\mathsfbi{R}^{(m)}\triangleq(i\omega\mathsfbi{I}-\mathsfbi{A}^{(m)})^{-1}$. In practice, the response, $\hat{\bm{q}}^{(m)}$, to a given forcing $\hat{\bm{f}}^{(m)}$ is computed by solving the linear system
\begin{align} \label{eq:lintosolve}
\mathsfbi{L}^{(m)}\hat{\bm{q}}^{(m)}=\hat{\bm{f}}^{(m)},
\end{align}
where $\mathsfbi{L}^{(m)}={\mathsfbi{R}^{(m)}}^{-1}=(i\omega\mathsfbi{I}-\mathsfbi{A}^{(m)})$ is a sparse linear operator. The resolvent code solves \eqref{eq:lintosolve} via LU decomposition using the PETSc library \citep{petsc-efficient}. Further details can be found in \cite{bugeat20193d}. 

The original code was written based on conservative variables, while the LES forcing database was generated using the primitive-like variable set, $\bm{q}=[\nu,\,\bm{u},\,p]$, as discussed in \S\ref{subsec:govern}, yielding a compatibility issue. To overcome this issue, a correction derived by \cite{karban_jfm_2020} is implemented.

 The sound level in the acoustic field is predicted by first computing the response to individual forcing realisations (or, instead, the RESPOD modes of the forcing) solving \eqref{eq:lintosolve} and then calculating the power spectral densities (PSD) based on these individual response realisations (or modes, instead).

\section{Identification of the acoustically efficient forcing components} \label{sec:ident}
{The goal of this study is to extract acoustically relevant forcing components that underpin noise-generation mechanisms in subsonic jets and to propose an empirical model for these components. Identification of the noise-generating part of the forcing prior to empirical modelling is crucial for the model to yield robust acoustic predictions. To achieve this, we proceed as follows.} Using the database M04Mc00, which contains both the state, $\bm{q}$, and the forcing, $\bm{f}$, we first conduct the analysis outlined in \S\ref{sec:math} based on RESPOD to identify the acoustically active forcing components. We limit the study to mechanisms associated with noise generation at low polar angles, which we refer to as downstream noise. We then discuss how to further decompose the low-rank forcing associated with the acoustic field to extract the part which satisfies the acoustic matching criterion \citep{fwilliams_rsta_1963,crighton_pas_1975}.
\subsection{Masking the forcing vector}
\begin{figure}
  \centerline{\resizebox{1\textwidth}{!}{\includegraphics{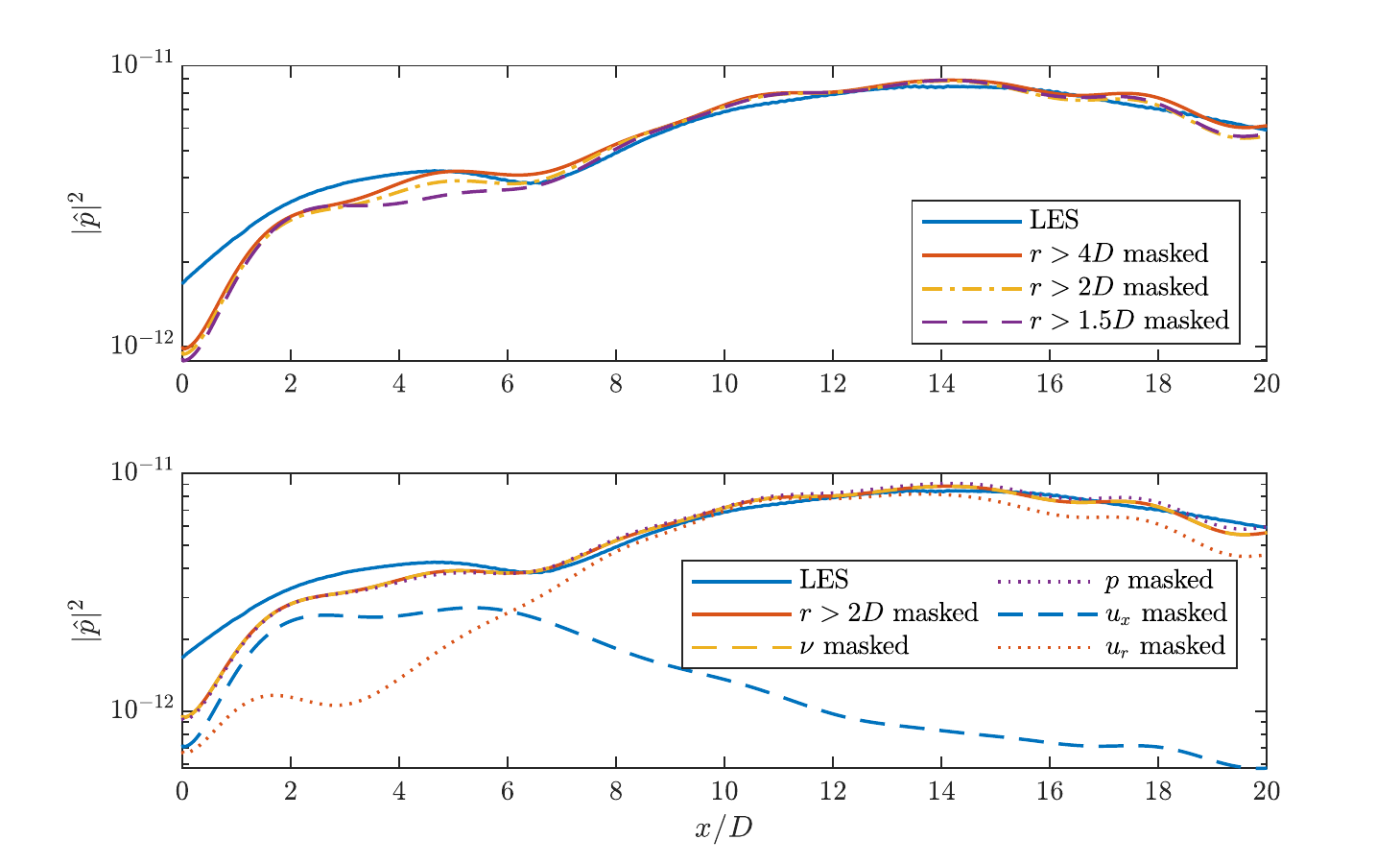}}}% Images in 100% size
  \caption{PSD of pressure predicted using resolvent analysis with masking applied in space (top) and in variables (bottom) in comparison to the LES data at $r=5D$ for the case M04Mc00 at $St=0.6$. }
\label{fig:pmask}
%\vspace{-15px}
\end{figure}

{Before performing a dedicated analysis to obtain a low-rank forcing model, we first reduce the number of the forcing terms to model.} This is achieved by applying spatial and componentwise masking in the forcing data to observe the effect of the masked regions/components on noise generation. This masking involves zeroing certain parts of the forcing vector using the matrix $\mathsfbi{B}$. Figure \ref{fig:pmask} shows the result of different masks in terms of the PSD of the acoustic pressure at $St=0.6$. Masking the forcing beyond $r>2D$ or $r>4D$ yields nearly identical results in the entire flow domain. Masking beyond $r>1.5D$ also yields nearly identical noise fields in the downstream region, $x>6D$, while a slight discrepancy is observed in the region $x/D\in[3,6]$. We therefore consider the forcing in the region $r<2D$ for the rest of the analysis. 

Componentwise masking of the forcing shows that the components $f_\nu$ and $f_p$, which correspond to the mass and energy equations, respectively, have negligible contribution to the acoustic field. Masking the component $f_{u_r}$, which is the forcing associated with the radial momentum equation, causes significant reduction in the sideline noise while not affecting the acoustic field in the downstream region. On the other hand, masking the component $f_{u_x}$, which is the forcing associated with the streamwise momentum, causes significant reduction in the downstream noise while having limited effect on the sideline noise. These results indicate that the forcing term $f_{u_x}$ in the region $r<2D$ is solely responsible for downstream noise generation, consistent with the observation of \cite{freund_jfm_2001} using Lighthill's analogy. {In what follows, focusing on the downstream noise generation only, we aim to identify the acoustically active subspace associated with this single forcing component.}

\subsection{Applying RESPOD to obtain low-rank forcing} \label{subsec:respoddata}
\begin{figure}
  \centerline{\resizebox{1\textwidth}{!}{\includegraphics{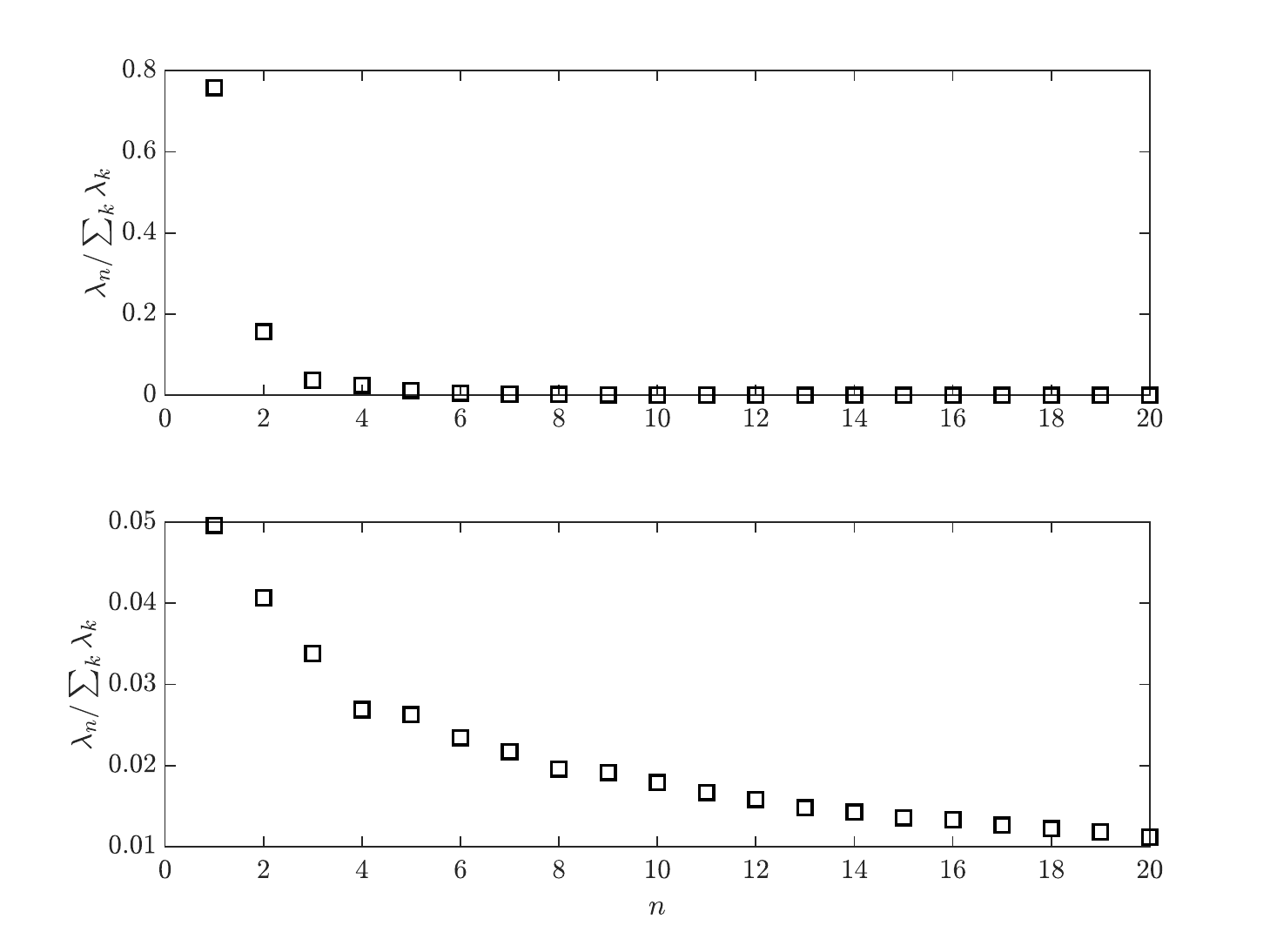}}}% Images in 100% size
\vspace{-10px}
  \caption{SPOD eigenvalues of the pressure in the acoustic field (top) and streamwise forcing, $f_{u_x}$, in the near field (bottom) for the case M04Mc00 at $St=0.6$. }
\label{fig:spodrespfrc}
\end{figure}

We now aim to obtain a low-rank representation of the subspace of the forcing associated with the most-energetic components of the acoustic field. To achieve this, we first compute the SPOD modes of the acoustic field, and we then use RESPOD to extract the associated forcing modes, as described in \S\ref{subsec:respod}. In figure \ref{fig:spodrespfrc}, the SPOD eigenvalues of the pressure in the downstream acoustic field, defined as $x/D,r/D\in[6,30]\times[4,6]$, and those of the forcing term $f_{u_x}$ in the turbulent region, defined as $x/D,r/D\in[0,30]\times[0,2]$, are shown for $St=0.6$. For the acoustic field, the leading SPOD eigenvalue corresponds to more than 75\% of the total acoustic energy. The sum of the first five SPOD eigenvalues corresponds to 99\% of the total acoustic energy, indicating a low-rank organisation in the acoustic field. For the forcing in the near field, on the other hand, the leading SPOD mode contains less than 6\% of the total energy in $f_{u_x}$. Around one hundred modes are required to capture 90\% of the total forcing energy, indicating an extremely high-rank structure.  As discussed earlier, this difference between the near-field turbulence forcing and the acoustic field is the crux of the jet-noise problem, which we aim to overcome by educing the small portion of the forcing responsible for the acoustic field.

\begin{figure}
  \centerline{\resizebox{1\textwidth}{!}{\includegraphics{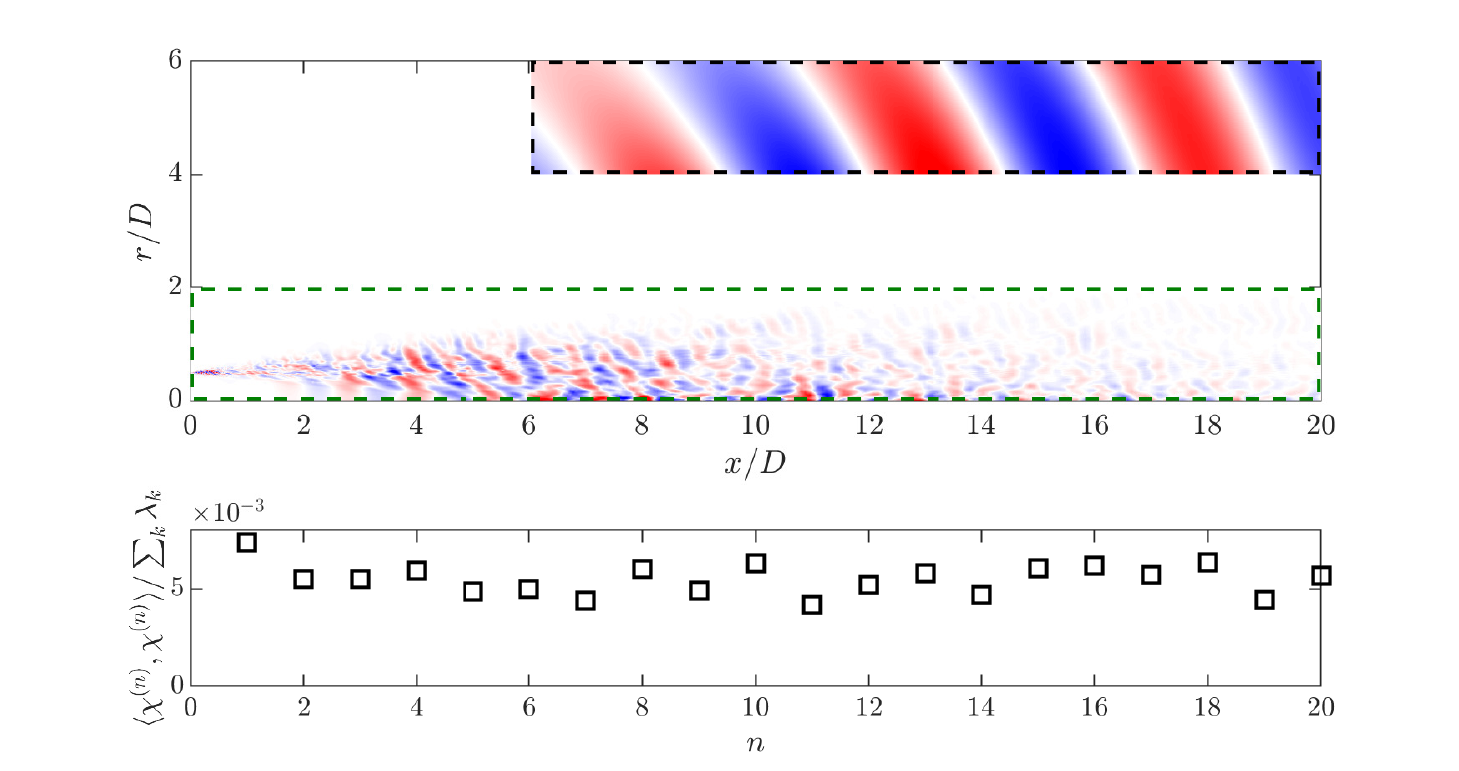}}}% Images in 100% size
\vspace{-10px}
  \caption{Optimal SPOD mode of acoustic pressure and the associated RESPOD mode of the forcing (top) together with the energy distribution in the first twenty RESPOD modes of the forcing (bottom) for the case M04Mc00 at $St=0.6$. The acoustic and forcing fields in the top plot are denoted by the black and green dashed boxes, respectively. }
\label{fig:spodmode}
\end{figure}

Using RESPOD, we extract from this high-rank forcing data a low-rank subspace that is correlated with the low-rank pressure structures observed in the acoustic field. In figure \ref{fig:spodmode}, we show the leading SPOD mode of pressure in the acoustic field and the associated RESPOD mode of the forcing in the near field, together with the energy distribution of the first twenty RESPOD forcing modes. {The leading SPOD mode takes the form of an acoustic wave propagating at some angle, while the associated RESPOD forcing mode contains a disorganised structure, which may imply underconvergence in the forcing mode, despite the very long time-series used for the analysis. The first RESPOD mode contains less than 0.8\% of the total forcing energy, but it is associated with the leading SPOD mode of the acoustic field, corresponding to 75\% of the total noise in the downstream region. This result shows the importance of applying such an identification prior to any modelling effort. Without this extraction of the low-rank forcing structure, a fitting function optimised using the forcing data will be affected by the existence of energetic structures that do not significantly contribute to sound generation.}

{No smooth trend is observed in the energy of the first twenty RESPOD forcing modes, contrary the SPOD modes in the acoustic field. As discussed in \cite{karban_jfm_2022}, RESPOD does not impose a strict filtering on the forcing to extract the \emph{active} structures that actually drive the acoustic field, but finds the correlated part which includes silent-but-correlated structures. Lack of a smooth trend in the energy of the RESPOD forcing modes implies underconvergence of these modes. Given that the active part in the RESPOD forcing modes are linearly related to the SPOD modes of the acoustic field, they should experience the same convergence rate. The underconvergence in the RESPOD forcing modes can then be attributed to the contribution of the silent-but-correlated structures, causing also the mode shape to be significantly less organised compared to the associated SPOD mode.} 
This underconvergence observed in the forcing modes does not pose a problem in the following analysis. The SPOD modes of the response and the RESPOD modes of the forcing are computed using Fourier realisations of response and forcing that exactly correspond to the same time window. In the ideal case of an error-free database, \eqref{eq:forceinfreq} is therefore satisfied for each pair of response-forcing realisations. So, no matter how underconverged the forcing data is, the structures generating the converged acoustic field are, by construction, ensured to be contained in the forcing mode seen in figure \ref{fig:spodmode}.

\begin{figure}
  \centerline{\resizebox{1\textwidth}{!}{\includegraphics{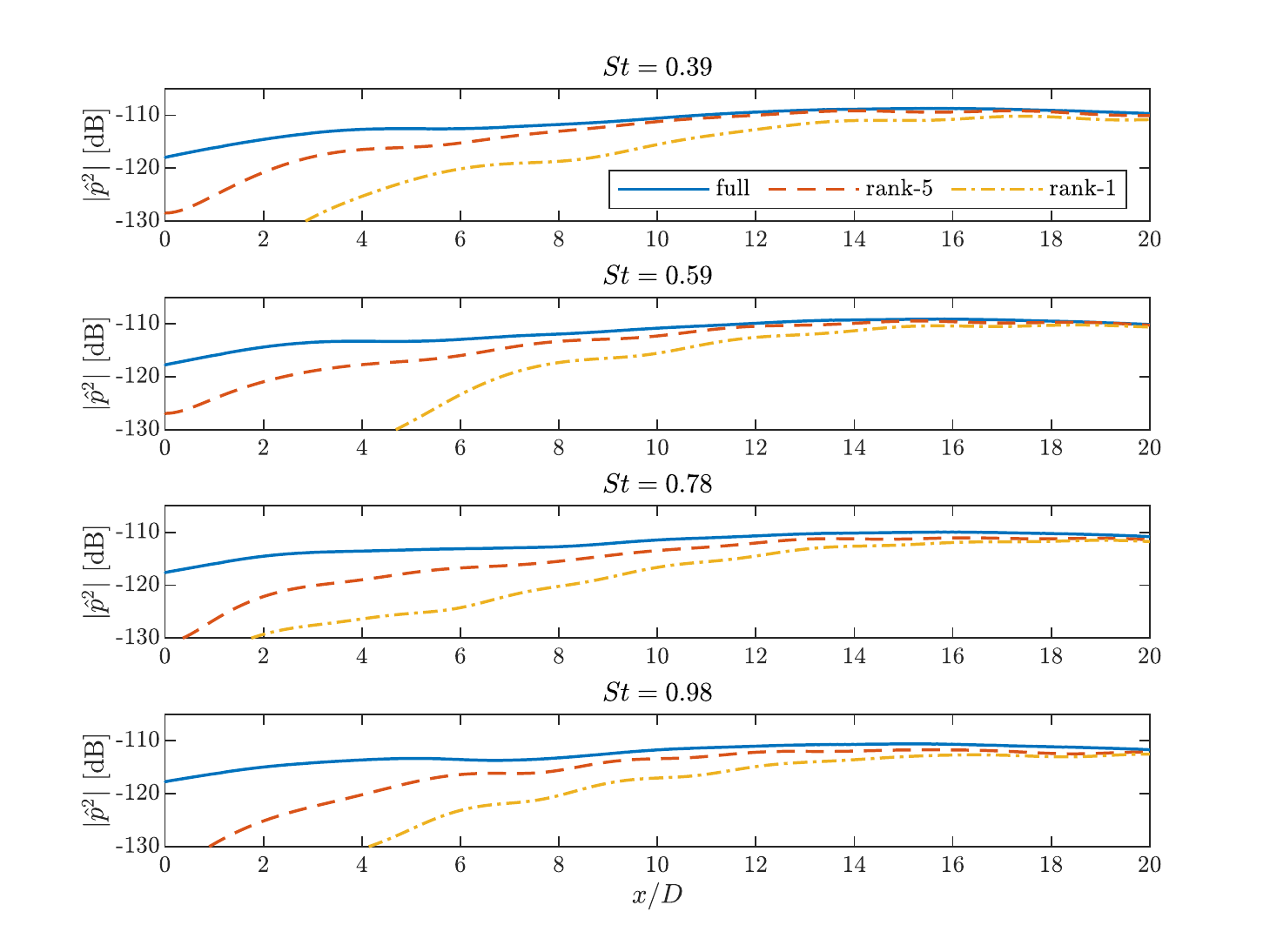}}}% Images in 100% size
\vspace{-15px}
  \caption{PSD of acoustic pressure generated using rank-5 and rank-1 forcing, respectively, obtained by RESPOD, in comparison to the acoustic field obtained from LES data (corresponding to full-rank forcing in the ideal case) at different frequencies ranging from $St=0.4$ to 1 (from top to bottom). }
\label{fig:fullvslowrank}
\end{figure}

As discussed in \S\ref{subsec:respod}, the first RESPOD forcing mode contains all structures correlated with the leading SPOD mode. This indicates that the remaining forcing that amounts to 99\% of the total forcing energy can generate only 25\% of the total acoustic energy. In figure \ref{fig:fullvslowrank}, we show a comparison of the true acoustic field and the acoustic fields obtained using rank-5 and rank-1 forcing truncations obtained using RESPOD for a number of frequencies, $St\in[0.4,1]$. The rank-5 forcing recovers nearly the entire acoustic field in the downstream region. The rank-1 forcing also recovers a significant portion of the downstream acoustic field. Although the acoustic field predicted by the rank-1 forcing should correspond to 75\% of the total acoustic energy, the actual prediction amounts to less than this ratio. This is due to the errors contained in the LES database, causing a loss in the correlation information between the response and the forcing. Despite all the limitations of the existing database as discussed in appendix \ref{subsec:error}, we see that it is still possible to define a rank-1 forcing which can generate most of the downstream noise. 

In what follows, we further decompose the rank-1 forcing obtained by RESPOD to extract the acoustically active forcing components which drive the leading SPOD mode of the acoustic pressure seen in figure \ref{fig:spodmode}.

\subsection{Isolating the radiating component of the low-rank forcing}
\begin{figure}
  \centerline{\resizebox{1\textwidth}{!}{\includegraphics{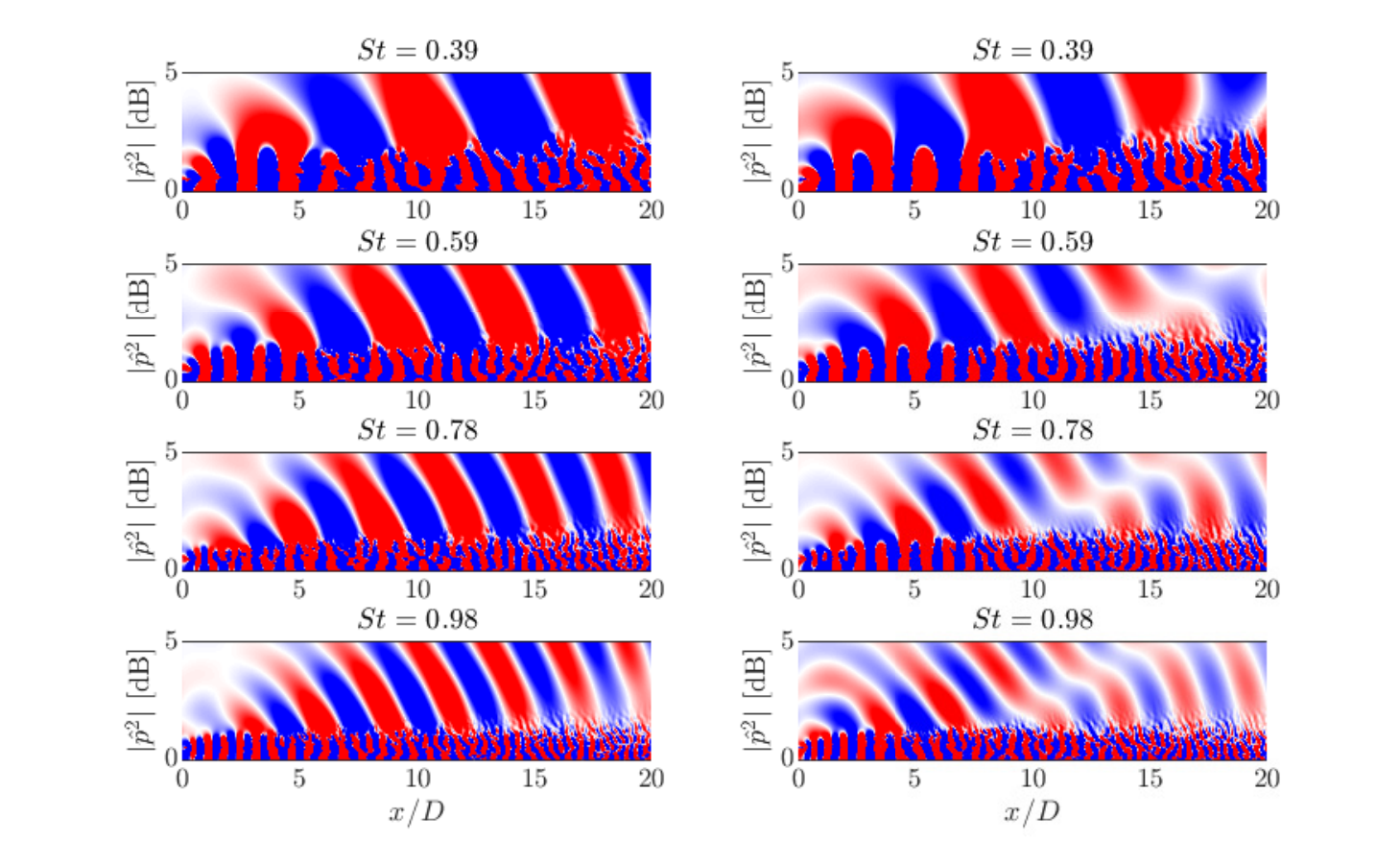}}}% Images in 100% size
\vspace{-15px}
  \caption{Real part of the pressure generated using first (left), and second (right) RESPOD mode of the forcing at different frequencies ranging from $St=0.4$ to 1 (from top to bottom). Color scale ranges between [$-1\times10^{-6},1\times10^{-6}$].}
\label{fig:respodpmode}
\end{figure}

In figure \ref{fig:respodpmode}, the acoustic fields generated by the first two RESPOD modes are shown for the case M04Mc00 at a number of frequencies, $St\in[0.4,1.0]$. {The first modes at all frequencies contain a single wave propagating at some angle with no jump in the phase. On the other hand, there exists a phase shift in the second modes that moves upstream with increasing frequency.} The phase shift appears in order to satisfy orthogonality between the first and the second modes, which is expected as RESPOD finds the forcing modes that generate the SPOD modes, which comprise an orthogonal basis. Note, however, that, no such orthogonality is ensured for the forcing modes. 

\begin{figure}
  \centerline{\resizebox{1\textwidth}{!}{\includegraphics{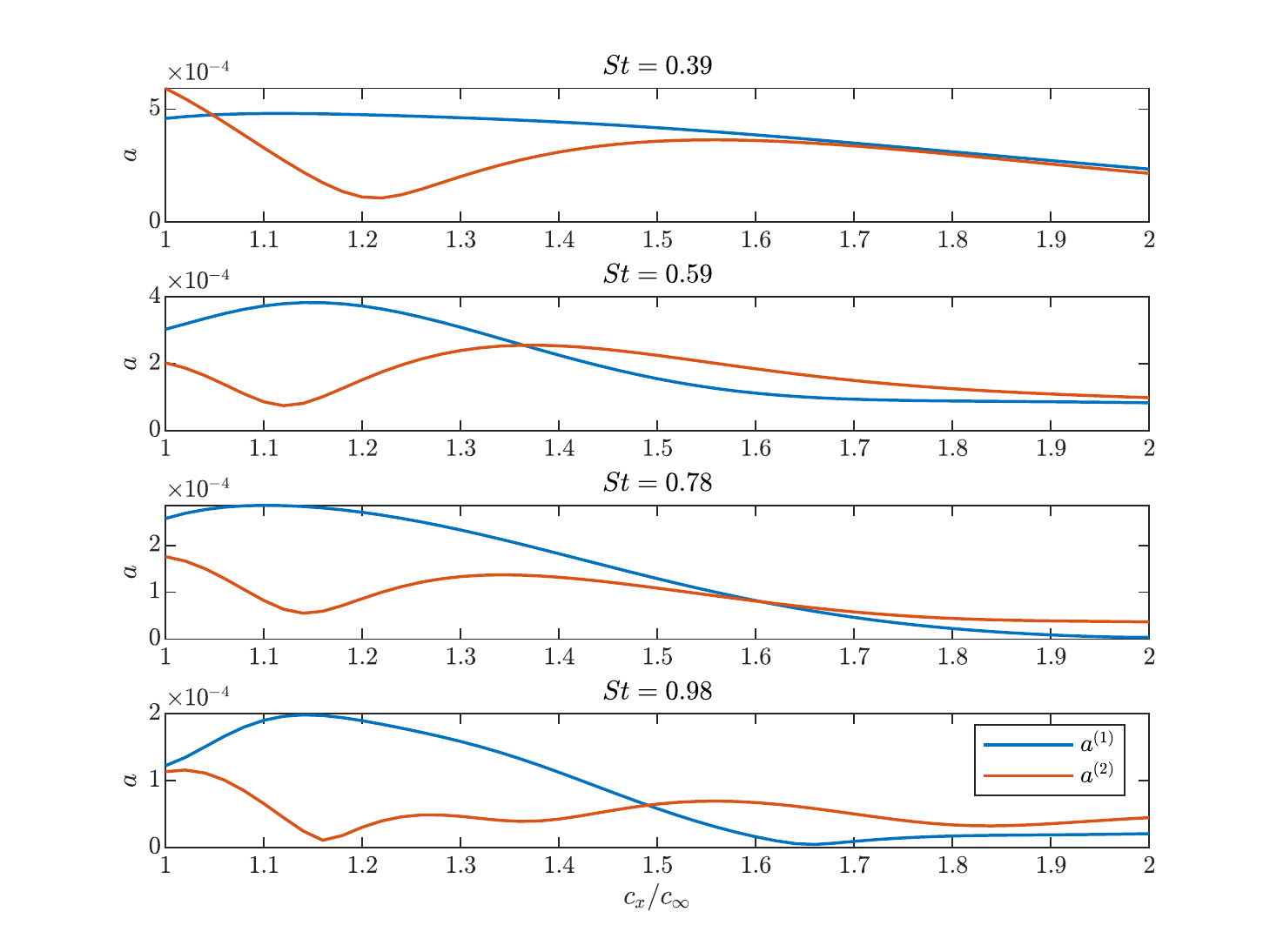}}}% Images in 100% size
\vspace{-15px}
  \caption{Projection of first and second RESPOD modes of the forcing, respectively, onto streamwise harmonic waves with supersonic phase speeds. Different frequencies ranging from $St=0.4$ to 1 are shown from top to bottom. }
\label{fig:projcoef}
\end{figure}

Looking at the acoustic field of the first RESPOD mode, it is apparent that the propagation angle is nearly constant,  around $30^\circ$ when measured from the downstream end, for all frequencies, reminiscent of a Mach-wave-like mechanism (c.f. \cite{tam_jfm_2008}). To explore this trend, we consider a wave in the streamwise direction defined by, $\exp(-ik_xx)$, where $k_x$ is the streamwise wavenumber associated with a phase speed
\begin{align}
c_x=\omega/k_x,
\end{align}
where $\omega=2\pi St$ is the angular frequency. For a Mach-wave-like mechanism, the phase speed is greater than the speed of sound, $c_\infty$, and the propagation angle is given by $\cos^{-1}(c_\infty/c_x)$ \citep{fwilliams_rsta_1963,crighton_pas_1975}. We project the first and the second RESPOD forcing modes onto this wave, varying in the phase speed over the range $[c_\infty,\,2c_\infty]$, yielding
\begin{align}
a^{(p)}(k_x,St)=\langle \bmit{\chi}^{(p)}(x,r,St),e^{-ik_xx}\rangle\triangleq\int_S\bmit{\chi}^{(p)}(x,r,St)e^{-ik_xx}dS,
\end{align}
where $p$ is the RESPOD mode number and $S$ is the 2D domain spanning the $x$ and $r$ directions. The results are shown in figure \ref{fig:projcoef}. It is seen that at all frequencies, the projection coefficient, $a^{(1)}$, peaks around 1.1-1.2$c_\infty$, which corresponds to an angle of $\sim30^\circ$, consistent with the propagation angle observed in the acoustic response field. The coefficient, $a^{(2)}$, on the other hand, has a dip around the same value at all frequencies, reminiscent of the orthogonality observed in the response modes of figure \ref{fig:respodpmode}.

\begin{figure}
  \centerline{\resizebox{1\textwidth}{!}{\includegraphics{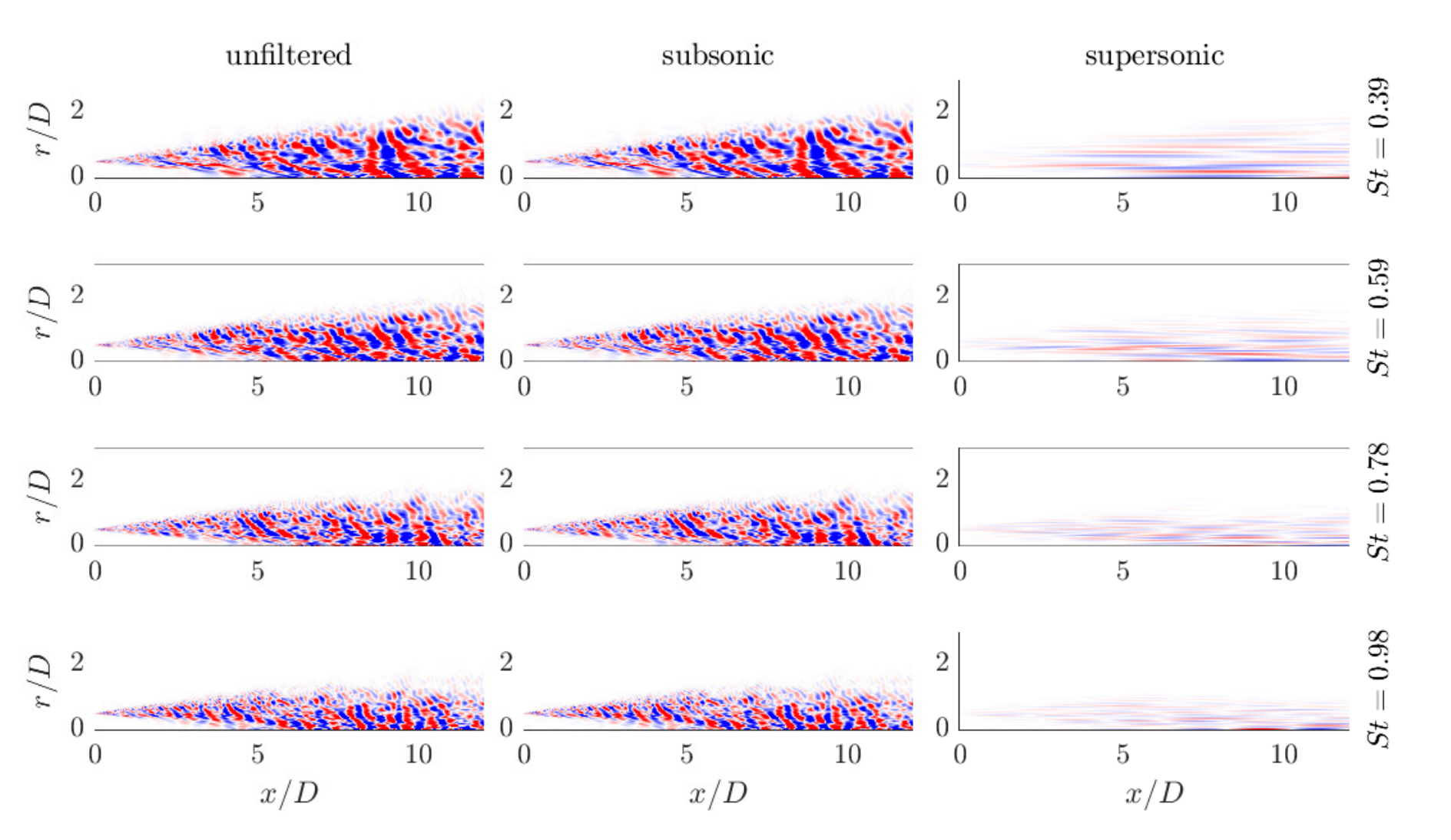}}}% Images in 100% size
\vspace{-15px}
  \caption{Real part of the first RESPOD mode of the forcing (left) compared to its subsonic (middle) and supersonic (right) parts. Different frequencies ranging from $St=0.4$ to 1 are shown from top to bottom. }
\label{fig:respodfilt}
\end{figure}

These results suggest that projection of the forcing onto supersonic waves is the relevant mechanism for generation of downstream noise, consistent with previous hypotheses and models \citep{freund_jfm_2001,cavalieri_jfm_2012,jordan_arm_2013,cavalieri_amr_2019}. To test this hypothesis, we define the following Fourier transform (FT) in the streamwise direction,
\begin{align}
\mathcal{F}_x(\mathsfbi{a})=\int_0^L\mathsfbi{a}e^{-ik_xx}dx,
\end{align}
where $L=30D$ is the domain length. Using this FT, we decompose the first RESPOD mode of the forcing, $\bmit{\chi}^{(1)}$, into two parts, $\bmit{\chi}^{(1-)}$ and $\bmit{\chi}^{(1+)}$, containing subsonic and supersonic components, respectively \citep{sinayoko_jfm_2011}. The resulting forcing fields are depicted in comparison to the original forcing mode in figure \ref{fig:respodfilt}. It is seen that most of the forcing energy is contained in the subsonic part of the mode, $\bmit{\chi}^{(1-)}$, making it indistinguishable from $\bmit{\chi}^{(1)}$. The supersonic component, $\bmit{\chi}^{(1+)}$, takes the form of a compilation of radially thin wavepackets with a disorganised radial phase structure. 

\begin{figure}
  \centerline{\resizebox{1\textwidth}{!}{\includegraphics{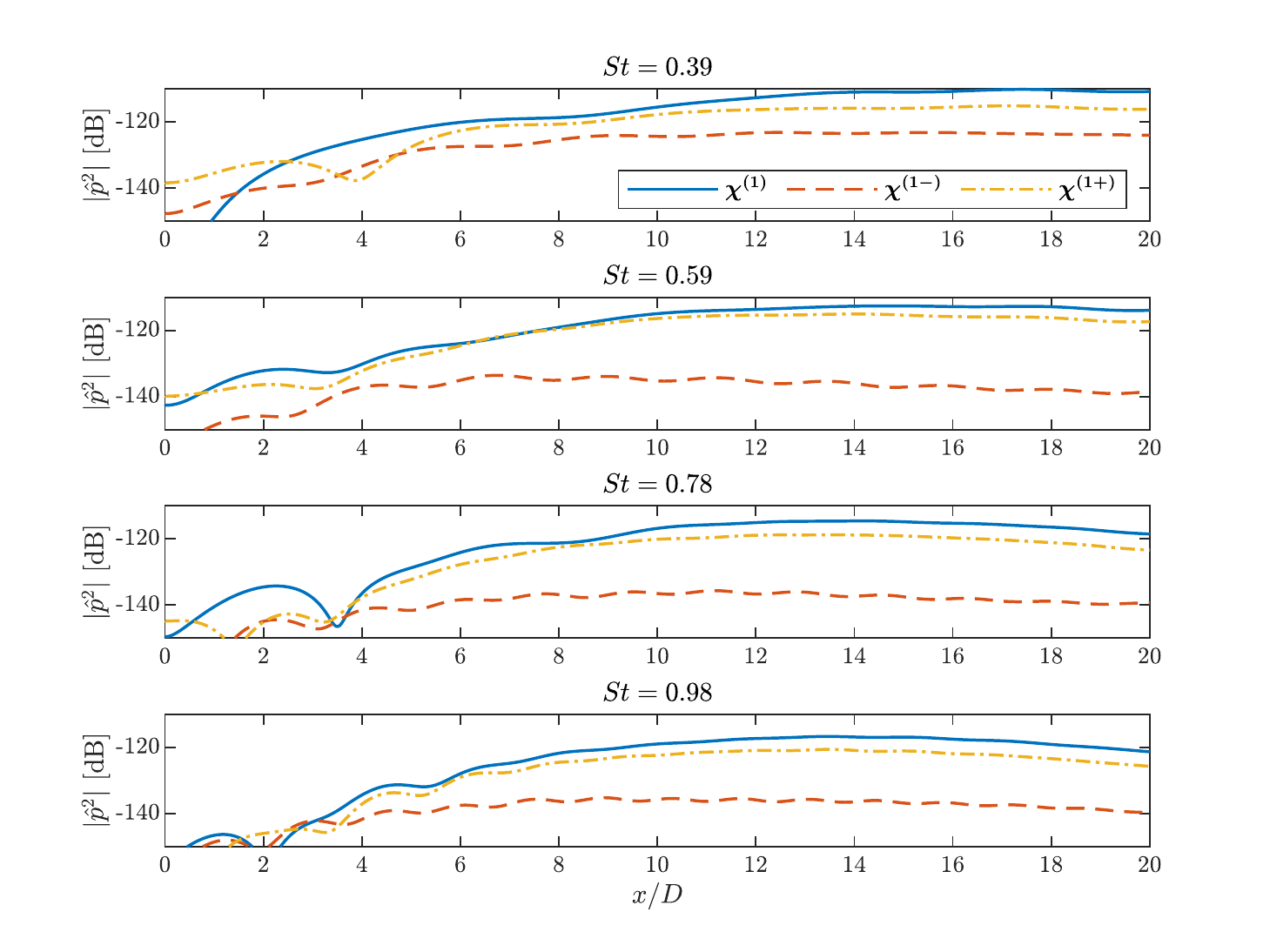}}}% Images in 100% size
\vspace{-15px}
  \caption{PSD of the acoustic pressure generated by the first RESPOD mode of the forcing (solid) compared to its subsonic (dashed) and supersonic (dash-dotted) parts. Different frequencies ranging from $St=0.4$ to 1 are shown from top to bottom. }
\label{fig:psubvssuper}
\end{figure}

\begin{figure}
  \centerline{\resizebox{1\textwidth}{!}{\includegraphics{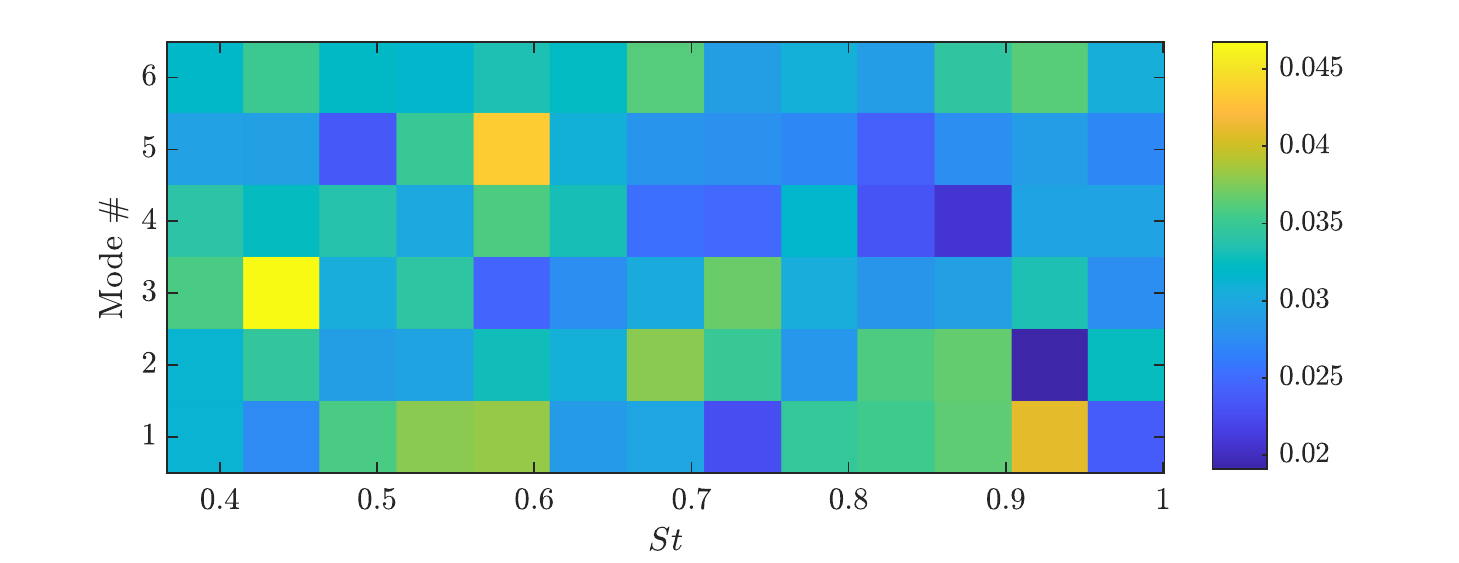}}}% Images in 100% size
%\vspace{-15px}
  \caption{Energy ratio of the supersonic part of the RESPOD mode of the forcing. }
\label{fig:filtegy}
\end{figure}

The acoustic response generated by these subsonic and supersonic modes, $\bmit{\chi}^{(1-)}$ and $\bmit{\chi}^{(1+)}$, respectively, are compared to the response of the entire first RESPOD mode of the forcing, $\bmit{\chi}^{(1)}$, in figure \ref{fig:psubvssuper} for a range of frequencies. It is clear that the supersonic modes underpin noise generation at all frequencies. Removing the supersonic components leads to more than an order-of-magnitude reduction in sound generation. The energy contained in the supersonic part of the RESPOD modes of the forcing is shown in figure \ref{fig:filtegy} for different mode numbers and frequencies. For all the modes and frequencies, the supersonic components contain less than 5\% of the mode energy. The first RESPOD mode of the forcing already contains less than $1\%$ of the total forcing energy, which means that the energy fraction of the supersonic part of the first RESPOD mode of the forcing, $\bmit{\chi}^{(1+)}$, with respect to the total forcing energy at the same frequency, is $\sim$0.04\%, while it generates $\sim75\%$ of the total acoustic energy in the downstream region for a frequency range $St=[0.4,1.0]$ at $M_j=0.4$.

\section{Empirical modelling of the acoustically efficient forcing component} \label{sec:model}
Having identified the acoustically efficient forcing components, we now characterise these modes and propose an empirical forcing model. In what follows, we first present a modelling strategy based on the supersonic part of the first RESPOD mode of the forcing for the case M04Mc00, which yields a fundamental form of the model equation. The model parameters are chosen such that they can be easily adapted to account for frequency, jet Mach number and flight effect. We then use the acoustic fields obtained from the LES to tune the parameters. Given the high energy contained in the optimal SPOD mode of the acoustic pressure in the downstream region, we focus on modelling the supersonic part of the first RESPOD mode of the forcing only; i.e., we construct a rank-1 model.

\subsection{Empirical source modelling for $M_j=0.4$}
As shown in figure \ref{fig:respodfilt}, the supersonic mode, $\bmit{\chi}^{(1+)}$, roughly follows the jet spreading and has the form of thin wavepackets elongated in the streamwise direction spanning most of the flow domain. Given the radial randomness of these thin wavepackets, a model for the $x$-$r$ structure is not feasible. We therefore make use of the characteristics of the modified resolvent operator, $\tilde{\mathsfbi{R}}$, which yields the acoustic pressure as the response thanks to the measurement matrix $\mathsfbi{C}$. 

\begin{figure}
  \centerline{\resizebox{1\textwidth}{!}{\includegraphics{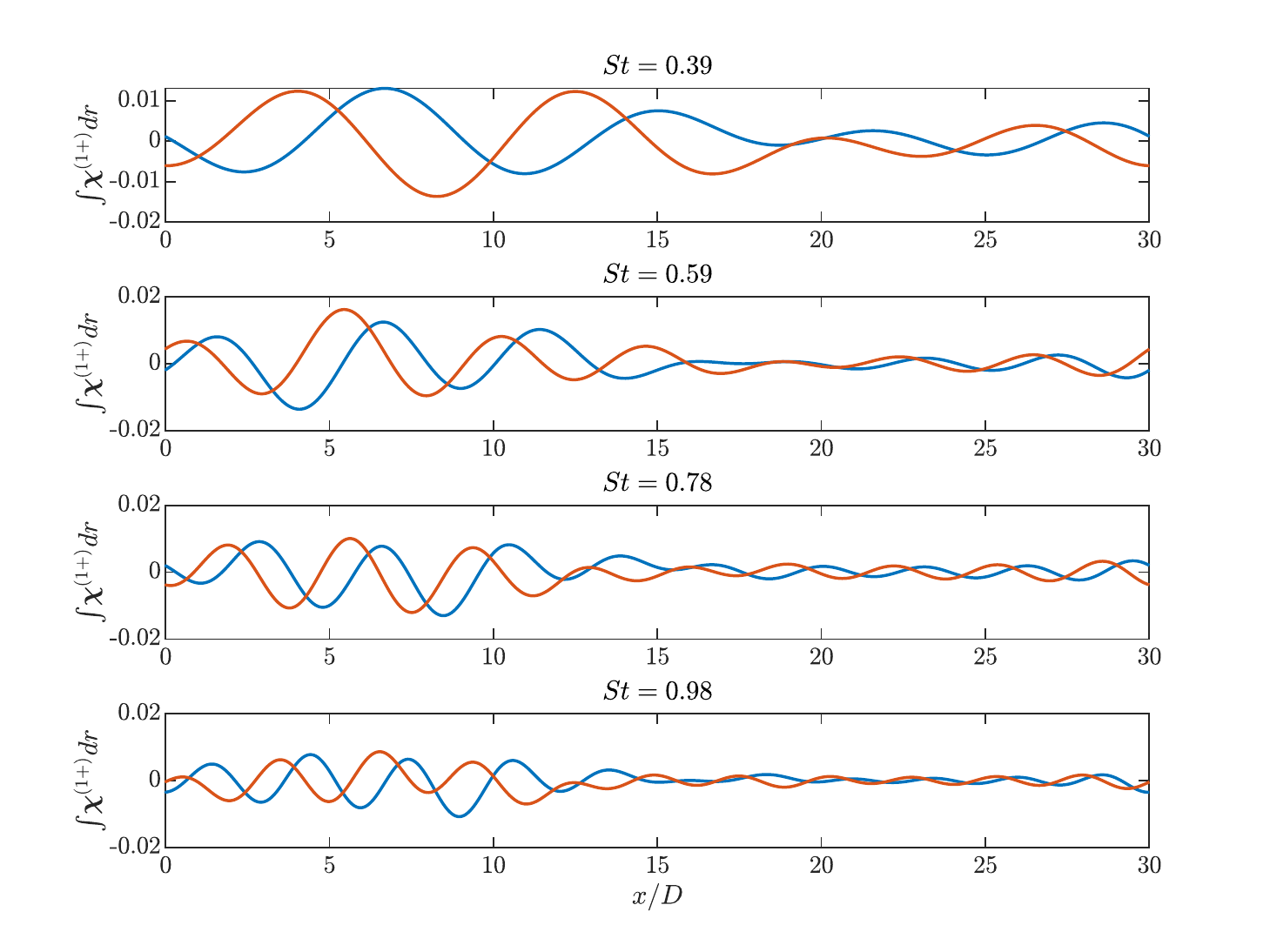}}}% Images in 100% size
\vspace{-15px}
  \caption{Real (blue) and imaginary (orange) parts of the supersonic part of the first RESPOD mode of the forcing integrated in the radial direction. Different frequencies ranging from $St=0.4$ to 1 are shown from top to bottom. }
\label{fig:linesrc}
\end{figure}

In resolvent analysis, a singular value decomposition (SVD) of the resolvent operator is used to identify mechanisms by which the output (acoustic pressure in our case) is driven by the input (forcing). The SVD is given as
\begin{align}
\tilde{\mathsfbi{R}}=\mathsfbi{U}\bmit{\Sigma}\mathsfbi{V}^H,
\end{align}
where $\mathsfbi{U}$ and $\mathsfbi{V}$ are the response and forcing modes, respectively, and $\bmit{\Sigma}$ denotes the resolvent gain.
{The RESPOD forcing mode is projected onto the forcing modes of the modified resolvent operator} and then amplified by the resolvent gains to generate the acoustic response whose spatial organisation is defined by the response modes. It was reported in \cite{jeun_pof_2016} and later in \cite{bugeat_prf_2023} that the forcing modes of the acoustic resolvent operator, $\tilde{\mathsfbi{R}}$, take the shape of streamwise supersonic waves with almost constant radial support. This indicates that the radially thin supersonic wavepackets observed in figure \ref{fig:respodfilt} are simply integrated in the radial direction as they are projected onto the forcing modes, which implies that the forcing modes of figure \ref{fig:respodfilt} can be replaced by a line source obtained by radial integration, justifying the use of line-source models in the literature \citep{michalke_dlr_1970,michel_aiaa_2009,
lesshafft_jfm_2010,cavalieri_jsv_2011,
cavalieri_jfm_2014,maia_rspa_2019,daSilva_jasa_2019}. The resulting line sources at different frequencies are shown in figure \ref{fig:linesrc} for the case M04Mc00. We observe wavepackets spanning $x/D=[0,20]$ with a dominant wavenumber at all frequencies. 

\begin{figure}
  \centerline{\resizebox{1\textwidth}{!}{\includegraphics{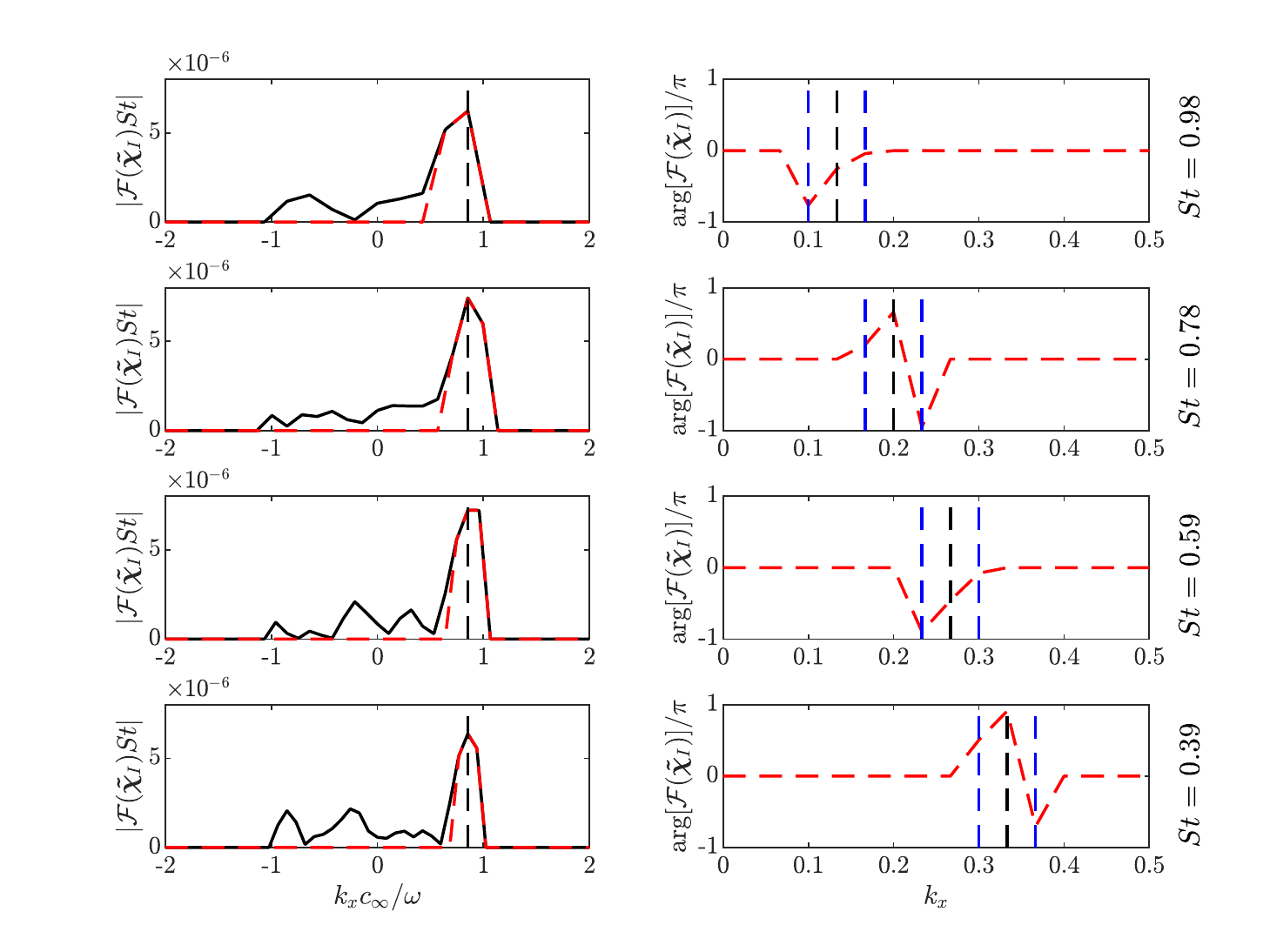}}}% Images in 100% size
\vspace{-15px}
  \caption{Amplitude (left) and phase (right) of the streamwise FT of the integrated line source. Red dashed line shows the isolated spectrum used for modelling. Vertical dashed black lines indicate the phase speed, $c_x=1.1722c_\infty$, and dashed blue lines mark the neighbouring wavenumbers to that. Different frequencies ranging from $St=0.4$ to 1 are shown from top to bottom. }
\label{fig:linesrcfft}
\end{figure}

\begin{figure}
  \centerline{\resizebox{1\textwidth}{!}{\includegraphics{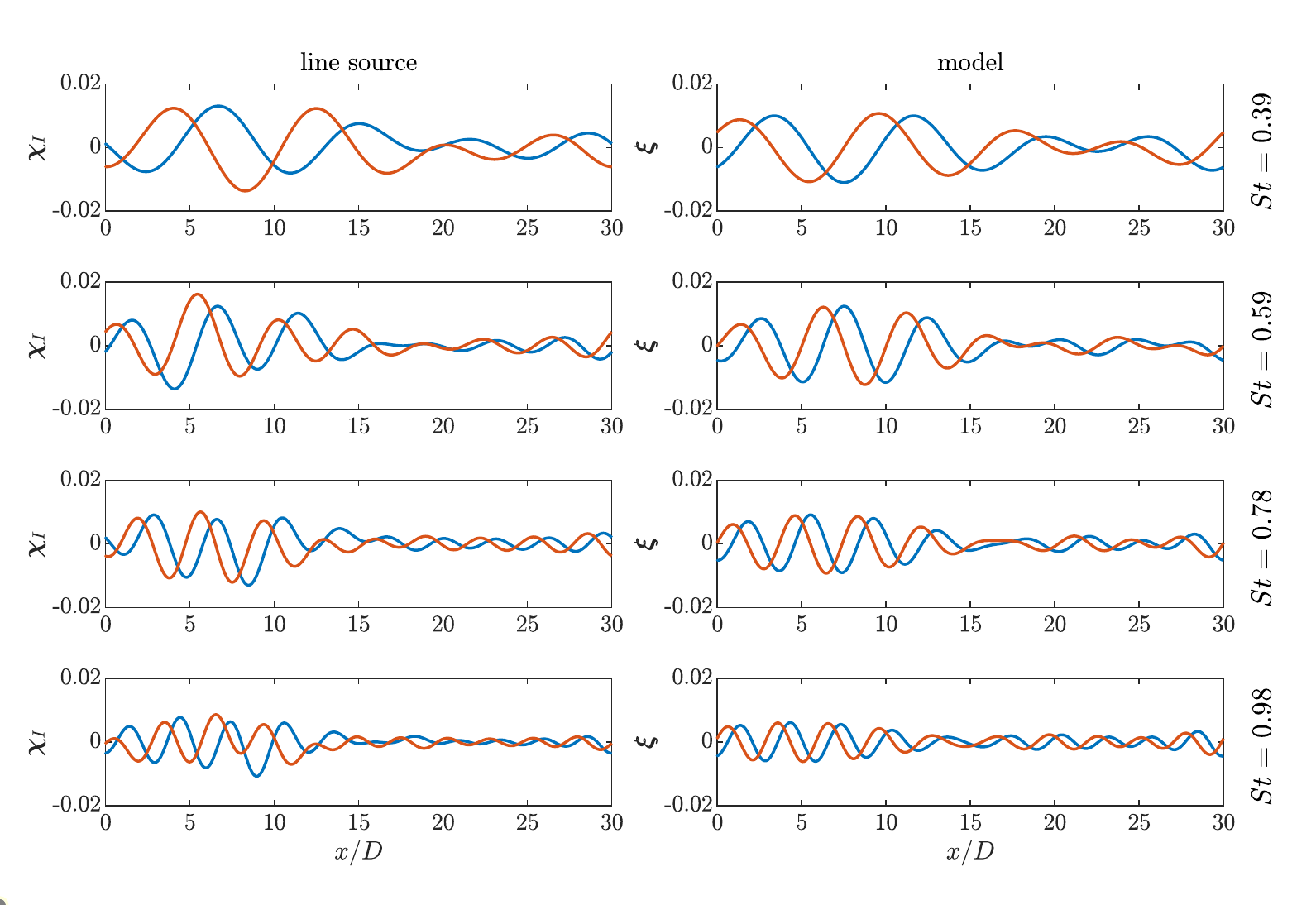}}}% Images in 100% size
\vspace{-15px}
  \caption{Real (blue) and imaginary (orange) parts of the line source (left) compared to the line-source model given by \eqref{eq:model04} (right). Different frequencies ranging from $St=0.4$ to 1 are shown from top to bottom.  }
\label{fig:linesrcmodel}
\end{figure}

To characterise these wavepackets, we perform a FT in the streamwise direction. The amplitude and phase of the Fourier coefficients are shown in figure \ref{fig:linesrcfft} for a range of frequencies. The wavenumber has been scaled by $c_\infty/\omega$ so that the abscissa in the figure is the inverse phase speed, with the range [-1,1] corresponding to the supersonic phase speeds. We see the same peak corresponding to a phase speed, $c_x=1.1722c_\infty$ at all frequencies, with energy contained in the immediate neighboring wavenumbers as well. The ordinate of the figure is scaled with $St$, and the peak has a nearly constant amplitude under this scaling. We ignore the rest of the spectrum and investigate the phase relation between this peak and the neighboring wavenumbers only, as shown in figure \ref{fig:linesrcfft}. There exists a phase difference of $\sim0.5\pi$ between the central and leftmost wavenumbers. The phase difference between the central and rightmost wavenumbers is either around $-1.7\pi$ or around $0.3\pi$, which correspond to the same phase shift as $0.3\pi=\textrm{mod}(-1.7\pi,2\pi)$. 

Given these observations, we need to find a model consisting of three wavenumbers, the wavenumber corresponding to the constant phase speed observed and the neighbouring ones, with some empirical phase and amplitude relations in between and a global amplitude which scales with $St$. We propose the following empirical model,
\begin{align}\label{eq:model04}
\mathcal{F}_x(\xi)=\frac{A}{St}\left(e^{i\pi k_x^p} + B\left(e^{\phi_1i\pi k_x^{p-}} + e^{\phi_2i\pi k_x^{p+}}\right) \right),
\end{align}
where $A=4.6\times 10^{-7}$, $B=0.8$, $\phi_1=0.5$, $\phi_2=-0.7$, $k_x^p=St/(\beta c_\infty)$ is the wavenumber corresponding to the peak observed in figure \ref{fig:linesrcfft} with $\beta=1.1722$, and $k_x^{p\pm}$ denotes the neighboring wavenumbers with $\Delta k_x=1/30$ where 30 is the domain length. The corresponding forcing model, $\xi$, can then be obtained by taking the inverse FT of \eqref{eq:model04} in $x$. Note that the neighboring wavenumbers provide the wavepacket amplitude envelope without having to define a Gaussian-like form. Choosing a different domain length would change $\Delta k_x$, and  therefore the resulting forcing model. But we anticipate that the results are not sensitive to this parameter, which will be justified later when showing the results for models at higher Mach numbers using the same value for $L$. Figure \ref{fig:linesrcmodel} shows a comparison of the forcing model obtained from \eqref{eq:model04} and the line source obtained from the LES data. The model has a similar spatial support and the same dominant wavelength corresponding to the line source. 

\begin{figure}
  \centerline{\resizebox{1\textwidth}{!}{\includegraphics{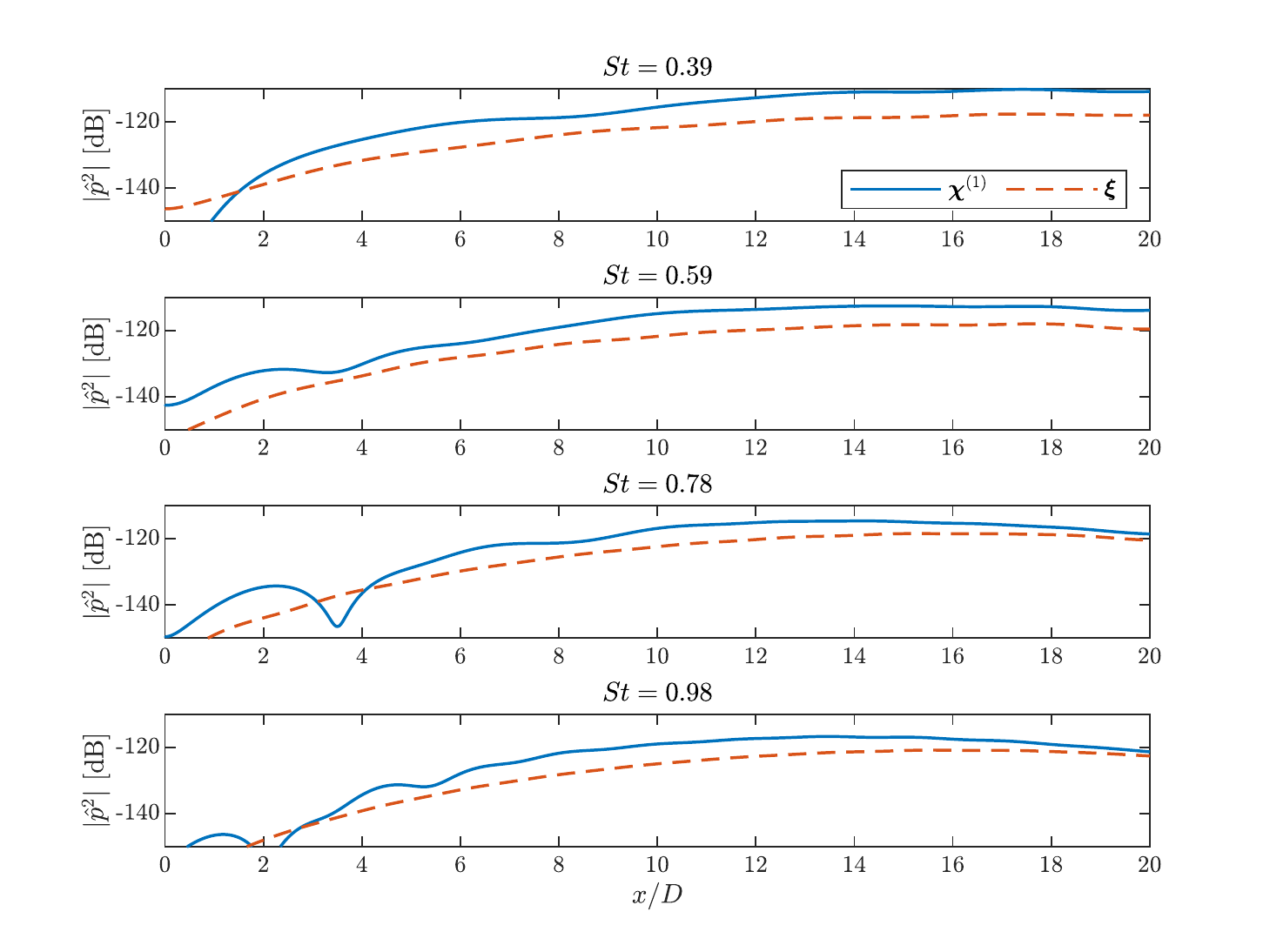}}}% Images in 100% size
\vspace{-15px}
  \caption{PSD of the acoustic pressure generated by the first RESPOD mode of the forcing (solid) compared to that of the line-source model (dashed) for the case M04Mc00. Different frequencies ranging from $St=0.4$ to 1 are shown from top to bottom.  }
\label{fig:pmodel}
\end{figure}

We now compare the acoustic response generated by this model to the response of the first RESPOD mode of the forcing, $\bmit{\chi}^{(1)}$, in figure \ref{fig:pmodel} for a range of frequencies. The model accurately predicts downstream noise generation for high frequencies while it yields an underprediction at lower frequencies. {The sound directivity is seen to be well captured at all frequencies, which implies that the underprediction at low frequencies can be fixed by adding a tuning parameter to the model. In the following section, we apply corrections to the model using acoustic data to improve predictions and to include Mach-number and flight effects.}

\subsection{Tuning the model using acoustic data from the LES}
{We discussed earlier that the errors in the LES database cause the correlation between the forcing and the acoustic field to be partially contaminated, yielding the response generated by the first RESPOD forcing mode, $\bmit{\chi}^{(1)}$, to globally underpredict the optimal SPOD mode of the acoustic field. As the forcing model in \eqref{eq:model04} is constructed based on the supersonic part of this forcing mode, the effect of the errors in the database is inherited in the model, $\xi$. To minimise this effect, we tune the model using the acoustic data obtained directly from the LES.} 

\begin{figure}
  \centerline{\resizebox{1\textwidth}{!}{\includegraphics{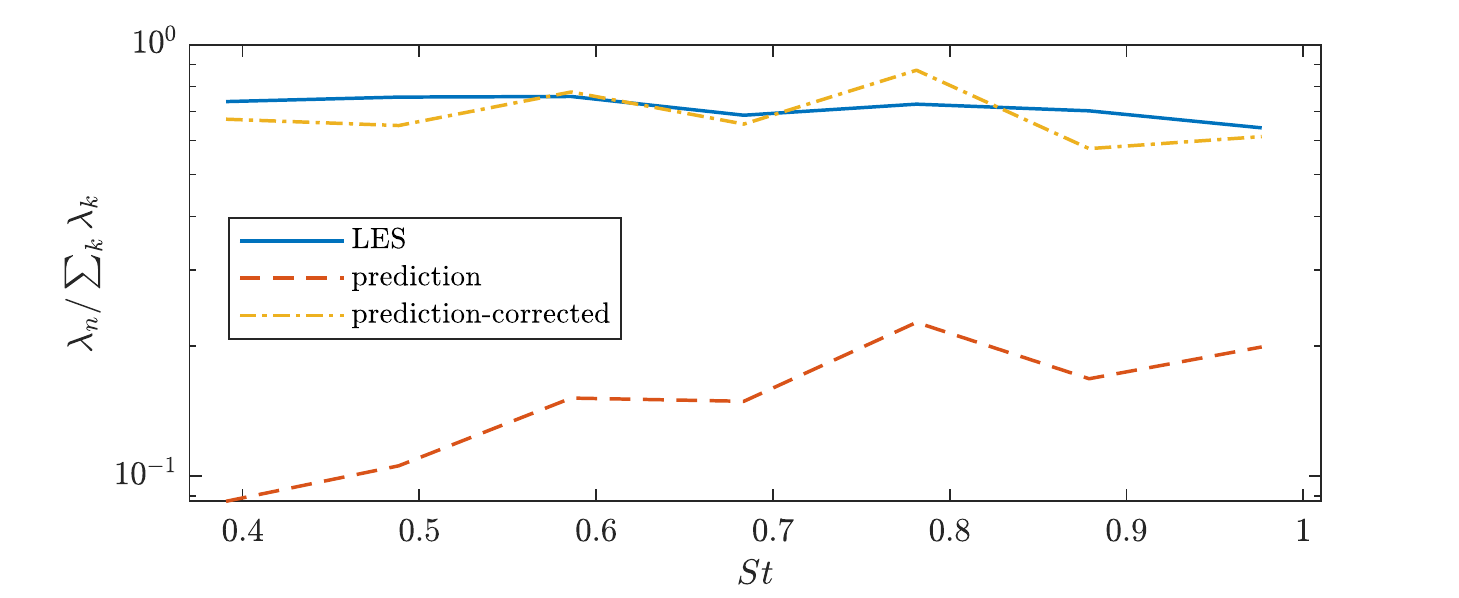}}}% Images in 100% size
\vspace{-5px}
  \caption{Energy ratio of the response generated by the line-source model (dashed) compared to that in the optimal SPOD mode of the acoustic pressure (solid). The energy ratio obtained using the corrected model is also shown (dash-dotted). Normalisations are done using the acoustic energy in the downstream region at each frequency.}
\label{fig:eratcomp}
\end{figure}

{We start tuning the model, $\xi$ by adding a scalar correction to better match the optimal SPOD mode of the acoustic field.} In figure \ref{fig:eratcomp}, the energy ratio of the response generated by the model, $\xi$, is compared with that of the optimal SPOD mode of the acoustic pressure as a function of frequency. Normalization is done using the total acoustic energy in the downstream region. It is seen that the underprediction of the model increases as the frequency decreases. A correction in the amplitude and changing the $St^{-1}$ scaling to $St^{-3/2}$ yields the corrected trend seen in figure \ref{fig:eratcomp}. 

We now test the ansatz \eqref{eq:model04} to see if it can capture the Mach-number effect. The phase difference between $k_x^p$ and $k_x^{p-}$ mainly determines the shape of the envelope of the wavepacket while the phase difference between $k_x^p$ and $k_x^{p+}$ determines its streamwise position. It was also found that the resulting acoustic field does not strongly depend on the value of $B$. Given these observations, we set $\beta$, $A$, $\phi_2$ as free parameters to tune, and we keep $\phi_1$ fixed to keep the wavepacket shape unchanged. 

Assuming that the parameter $\beta$ varies linearly with the jet Mach number $M_j$ and matching the observed value at $M_j = 0.4$ and the observed phase velocity at $M_j = 0.9$ yields the expression
\begin{align} \label{eq:betamj}
\beta=0.7722 + M_j.
\end{align}
For $M_j=0.9$, this results in a propagation angle of $53.3^\circ$, very close to the value observed by \cite{bugeat_prf_2023}. 

\begin{figure}
  \centerline{\resizebox{1\textwidth}{!}{\includegraphics{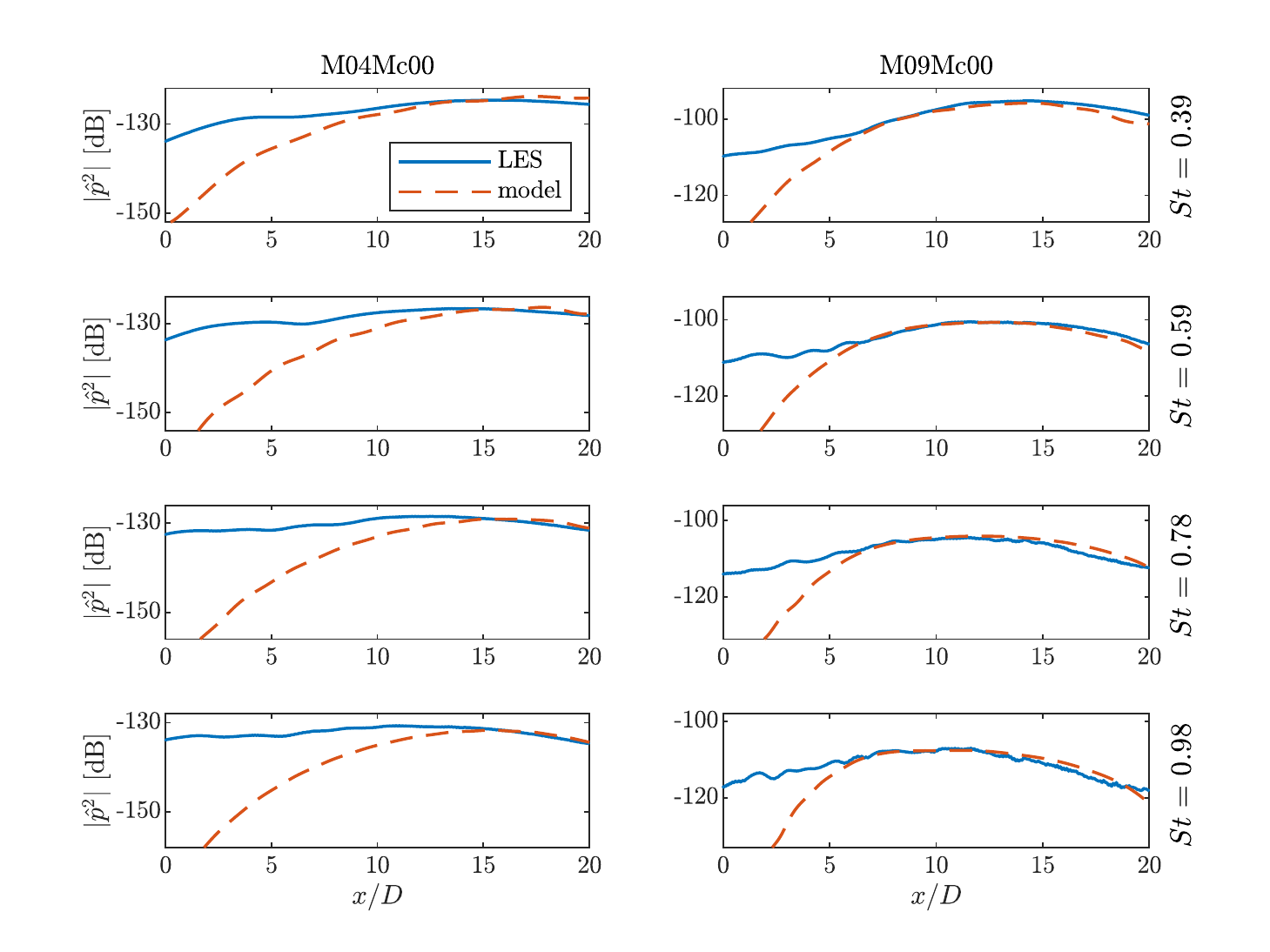}}}% Images in 100% size
\vspace{-15px}
  \caption{PSD of the acoustic pressure obtained using line-source model \eqref{eq:model09} to that extracted from the LES for the cases M04Mc00 (left) and M09Mc00 (right). Different frequencies ranging from $St=0.4$ to 1 are shown from top to bottom.}
\label{fig:respredict}
\end{figure}

To set the amplitude, $A$, and the phase constant, $\phi_2$, we performed tests to find the parameters that best match the acoustic field in the case M09Mc00. We finally obtained the empirical relations
\begin{align} \label{eq:amp}
A&=3.22\times10^{-6}M_j^{7/2},\\
\phi_2&=0.1-St. \label{eq:phi2}
\end{align}
The resulting model equation with these corrections reads
\begin{align}\label{eq:model09}
\mathcal{F}_x(\xi)=3.15\times10^{-6}\sqrt{\frac{M_j^7}{St^3}}\left(e^{i\pi k_x^p} + B\left(e^{0.5i\pi k_x^{p-}} + e^{(0.1-St)i\pi k_x^{p+}}\right) \right),
\end{align}
where
\begin{align}
k_x^p=\frac{St}{(0.7722+M_j)c_\infty},
\end{align} 
and $k_x^{p\pm}$ denotes once again the neighboring wavenumbers with $\Delta k_x=1/30$. The resulting acoustic field is shown in figure \ref{fig:respredict} for M04Mc00 and M09Mc00 and compared to the corresponding LES data. It is seen that the model given in \eqref{eq:model09} yields a prediction that well matches the LES data in the downstream region at all frequencies. {The acoustic response does not noticeably differ for the case M04Mc00 whether one uses a constant or linearly varying value for the phase, $\phi_2$.}

\begin{figure}
  \centerline{\resizebox{1\textwidth}{!}{\includegraphics{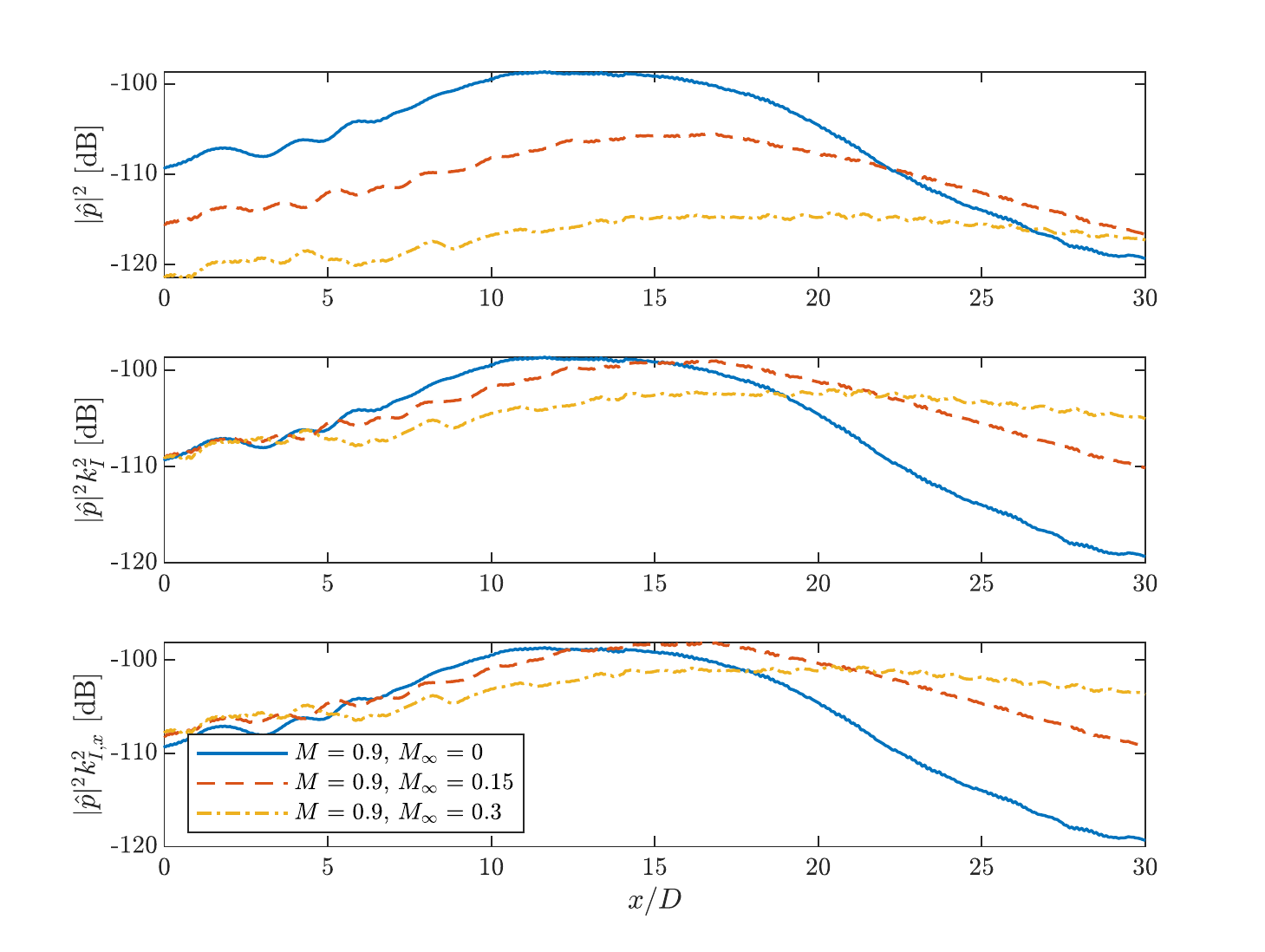}}}% Images in 100% size
\vspace{-5px}
  \caption{PSD of the acoustic pressure for the cases M09Mc00 (solid), M09Mc15 (dashed) and M09Mc30 (dash-dotted) with no scaling (top), $k_I^2$ scaling (center) and $k_{I,x}^2$ scaling (bottom) at $St=0.6$.}
\label{fig:coflow}
\end{figure}

\begin{figure}
  \centerline{\resizebox{1\textwidth}{!}{\includegraphics{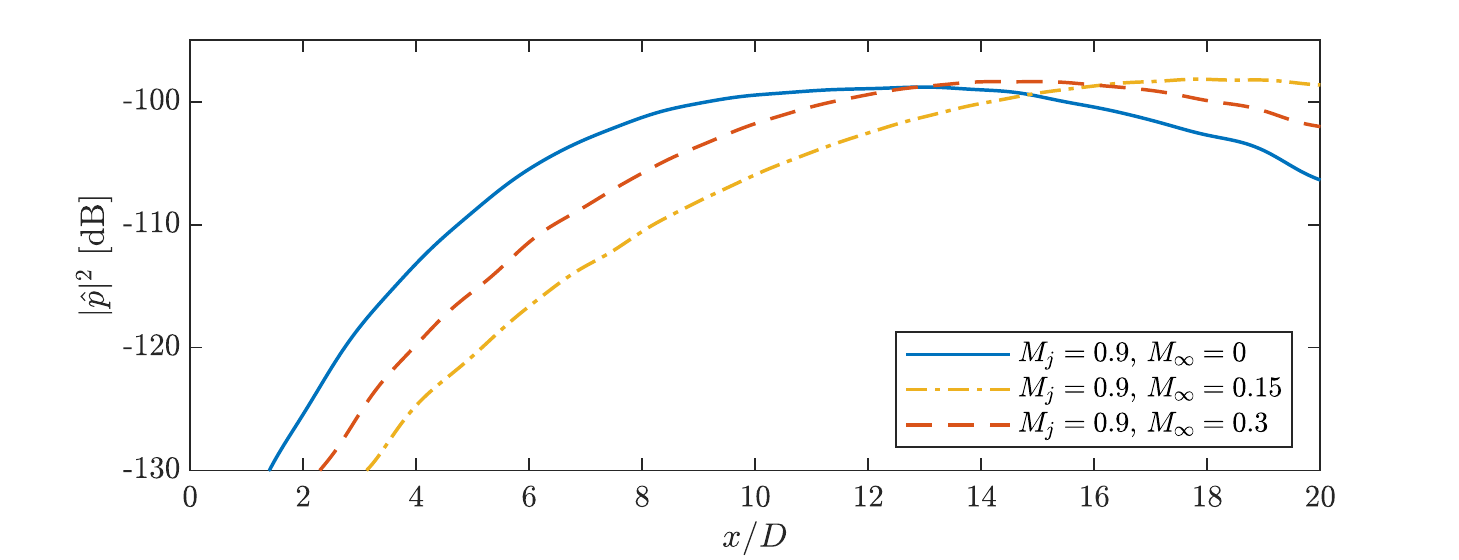}}}% Images in 100% size
\vspace{-0px}
  \caption{PSD of the acoustic pressure obtained using the line-source model given in \eqref{eq:model09} for the cases M09Mc00 (solid), M09Mc15 (dashed) and M09Mc30 (dash-dotted) at $St=0.6$.}
\label{fig:coflowmodel1}
\end{figure}

\begin{figure}
  \centerline{\resizebox{1\textwidth}{!}{\includegraphics{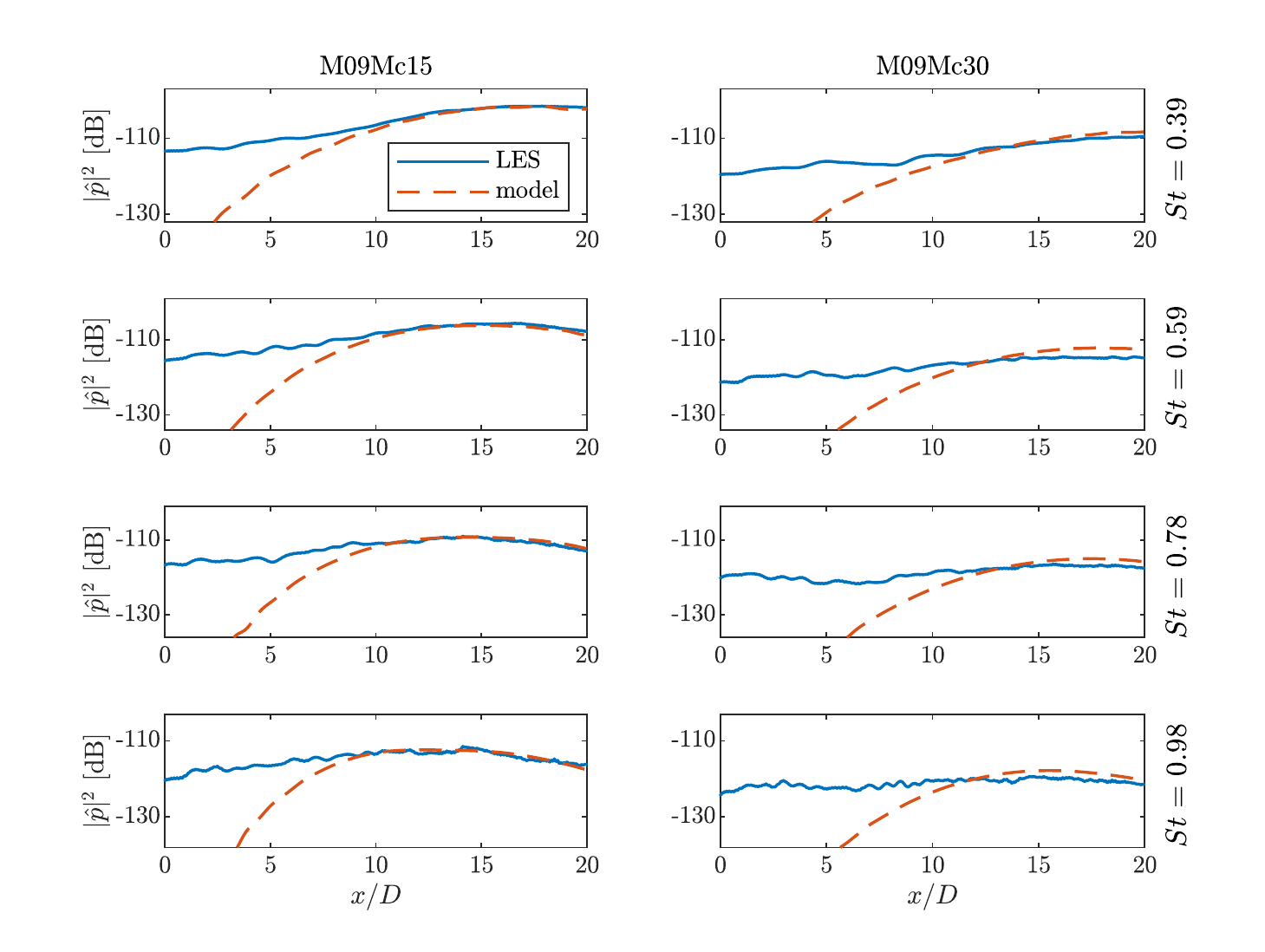}}}% Images in 100% size
\vspace{-5px}
  \caption{PSD of the acoustic pressure obtained using the line-source model given in \eqref{eq:modelfinal} (dashed) compared to the LES data (solid) for the cases M09Mc15 (left) and M09Mc30 (right). Different frequencies ranging from $St=0.4$ to 1 are shown from top to bottom.}
\label{fig:coflowmodel2}
\end{figure}

Finally, we extend the empirical model given in \eqref{eq:model09} to take into account the flight effect. It is known that the effect of flight is to suppress noise, largely due to the suppression of turbulence in the shear layer \citep{maia_arxiv_2022}. We compare the noise generated in the cases M09Mc00, M09Mc15 and M09Mc30, respectively, in figure \ref{fig:coflow} with two different scalings. Defining,
\begin{align}
k_I(M_j,M_\infty)=\int_SK(M_j,M_\infty)dS,
\end{align}
where $K(M_j,M_\infty)$ denotes the turbulent kinetic energy  as a function of the jet Mach number, $M_j$, and the flight Mach number, $M_\infty$, and using the scaling $k_I^2$ causes the sidestream noise in the three cases to collapse on top of each other. The dominant terms in the forcing have the form $\bm{u}\cdot\nabla\bm{u}$, which has the same dimension with turbulent kinetic energy differentiated in space. Inspired from this, we defined a scaling factor using $\partial_xK$ as 
\begin{align}
k_{I,x}(M_j,M_\infty)=\int_S\partial_xK(M_j,M_\infty)dS.
\end{align}
The scaling $k_{I,x}^2$ yields a slightly improved match in the peak noise level of the three cases. To determine if this scaling should be directly adopted in the forcing model, one needs to understand the effect of the free stream on the efficiency of the noise generation mechanisms embedded in the resolvent operator. For this, we performed a resolvent-based noise prediction using the same forcing in the three cases at $M_j=0.9$. The resulting acoustic fields are shown in figure \ref{fig:coflowmodel1}, where it is seen that the change in the mean flow does not strongly affect noise level, but causes a change in the directivity in a similar fashion as observed in the LES data seen in figure \ref{fig:coflow}. This suggests that one may use the same mathematical form for the source model for jets with or without flight effect, applying an amplitude correction. We adopt $k_{I,x}^2$ scaling for the empirical model as the peak noise, which occurs in the downstream region, is more relevant for the present study. Besides the amplitude correction, comparison of the model prediction with the LES data with flight stream effect yielded that the phase constant, $\phi_2$, is to be updated as
\begin{align}
\phi_2=0.1 - St\frac{M_j-M_\infty}{M_j},
\end{align}
resulting in the final equation for the jet-noise source model,
\begin{align}\label{eq:modelfinal}
\mathcal{F}_x(\xi)=3.15\times10^{-6}\,\Gamma\sqrt{\frac{M_j^7}{St^3}}\Biggl(e^{i\pi k_x^p} + B\left(e^{0.5i\pi k_x^{p-}} + e^{(0.1-St(M_j-M_\infty)/M_j)i\pi k_x^{p+}}\right) \Biggr),
\end{align}
where
\begin{align}
\Gamma\triangleq\frac{k_{I,x}(M_j,M_\infty)}{k_{I,x}(M_j,0)}.
\end{align}
The resulting acoustic response of the model forcing obtained by taking the inverse FT of \eqref{eq:modelfinal} is compared to the acoustic fields coming from the LES data in the cases M09Mc15 and M09Mc30, respectively, in figure \ref{fig:coflowmodel2}. The error at all the frequencies remains within 2dB for the downstream region. 

\subsection{Blind testing the model under different operating conditions}
\begin{figure}
  \centerline{\resizebox{1\textwidth}{!}{\includegraphics{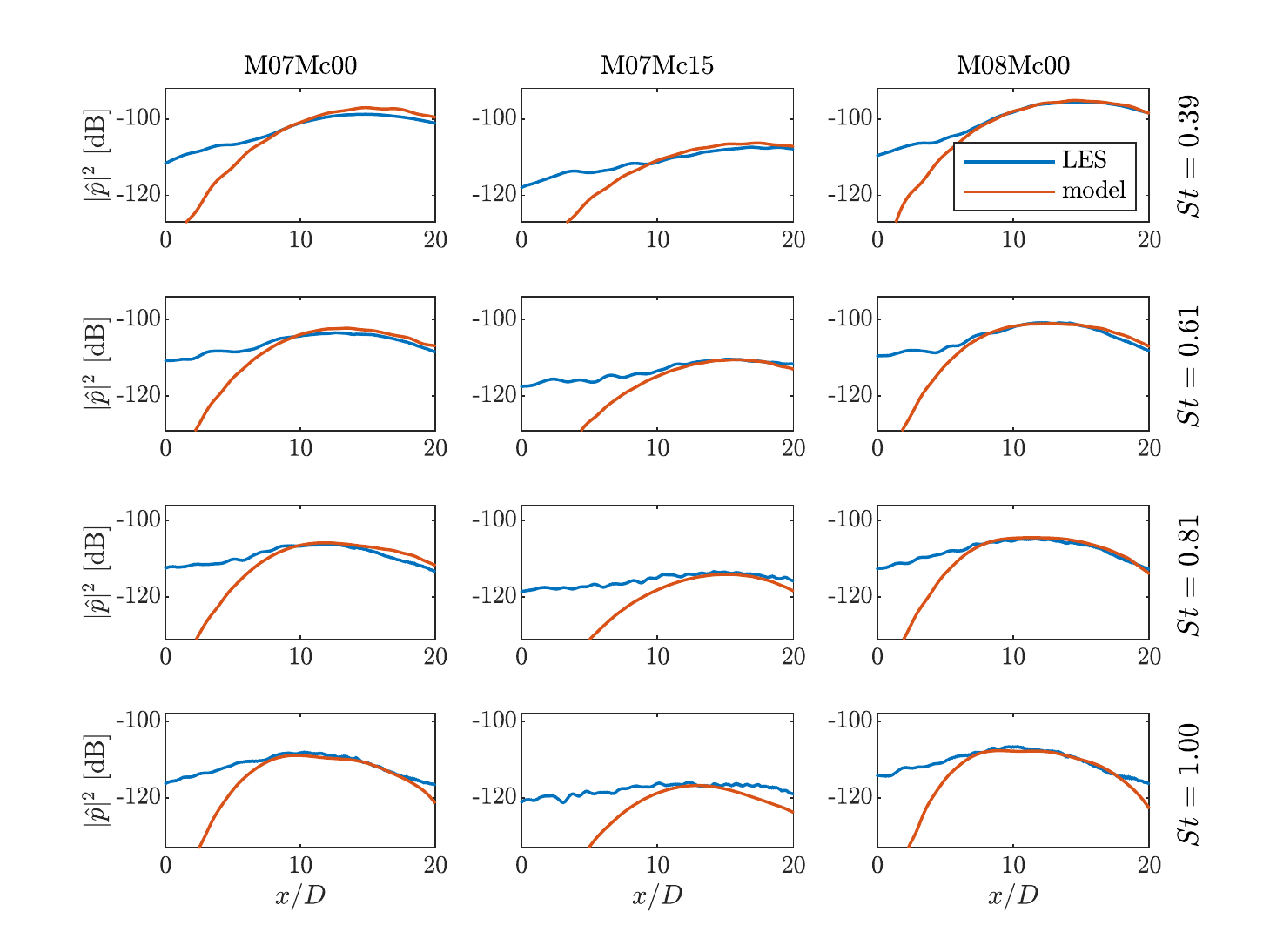}}}% Images in 100% size
\vspace{-5px}
  \caption{PSD of the acoustic pressure obtained using the line-source model given in  \eqref{eq:modelfinal} (dashed) compared to the LES data (solid) for the cases M07Mc00 (left), M07Mc15 (center) and M08Mc00 (right). Different frequencies ranging from $St=0.4$ to 1 are shown from top to bottom.}
\label{fig:coflowmodel3}
\end{figure}

{To test the validity of the model given in \eqref{eq:modelfinal}, we use three LES cases that were not used in its development: M07Mc00, M07Mc015 and M08Mc00 (see table \ref{tab:les} for details). The predictions are compared against the LES data in figure \ref{fig:coflowmodel3}. The acoustic response generated by the forcing model predicts the downstream acoustic field with a ~1dB accuracy within the region $x/D=[10,20]$ for the static jets at $M_j=0.7$ and 0.8, except for a sharper decay observed at $St=1$ beyond $x=19D$ for both cases. For the case with flight effect, the peak noise level is predicted accurately at all the frequencies. The accuracy of the prediction is within 2 dB for the region $x/D=[10,20]$ at $St=0.4$ and 0.6 and for the region $x/D=[12,20]$ at $St=.8$. A sharper decay is observed for $St=1$ at $x=14D$. These results show that the proposed forcing model is capable of predicting jet noise within the Mach number and Strouhal number ranges $M_j=[0.4,0.9]$ and $St=[0.4,1]$, respectively. The suppression of jet noise due to flight effect is also well captured
for the regime $M_\infty/M_j<0.33$. Beyond these limits, the validity of the model remains to be tested.}

\begin{figure}
  \centerline{\resizebox{1\textwidth}{!}{\includegraphics{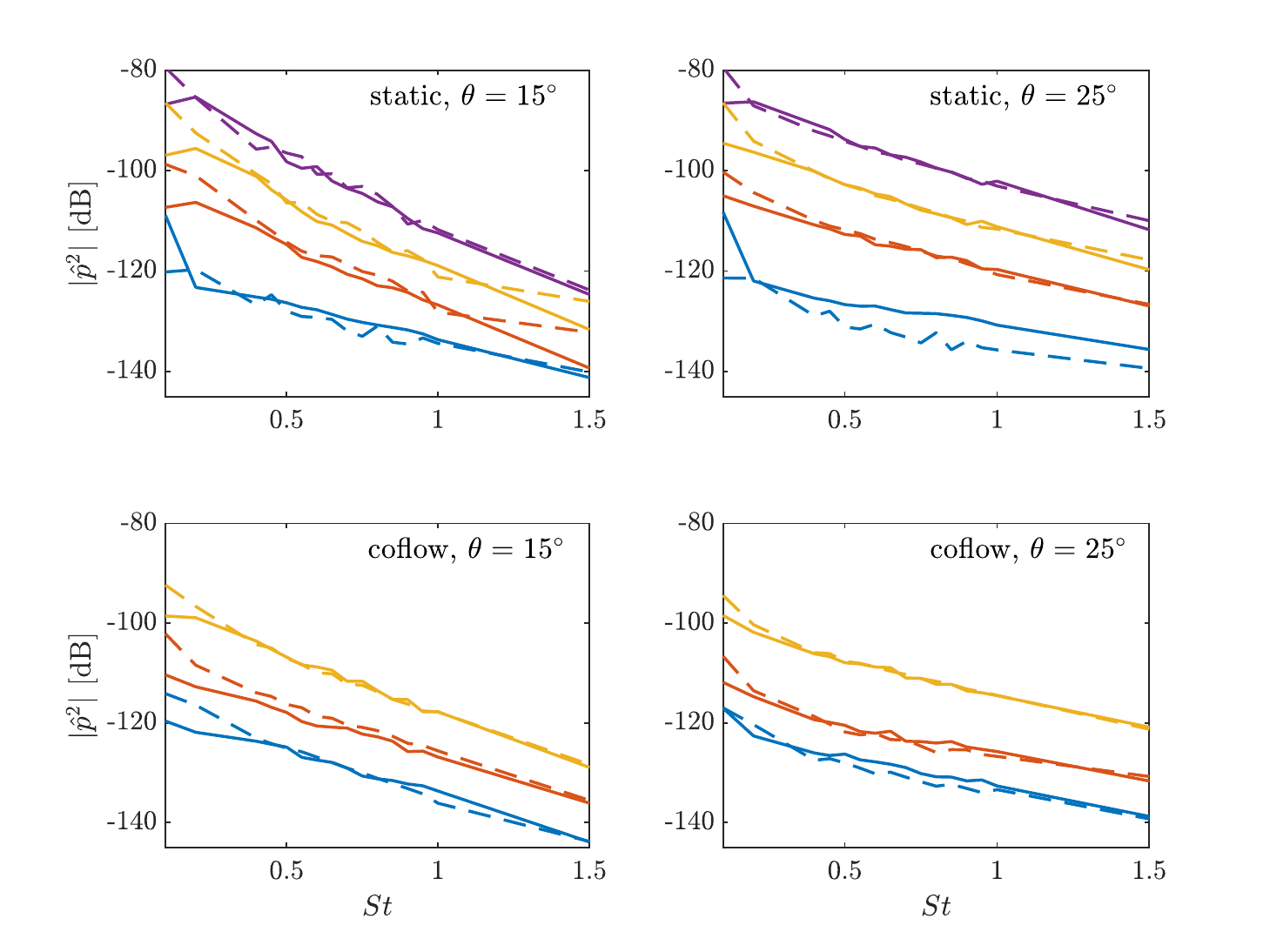}}}% Images in 100% size
\vspace{-5px}
  \caption{PSD of the acoustic pressure obtained from LES (solid) and predicted by the line-source model (dashed) for static jet cases (top) and cases with flight stream (bottom) at two different propagation angles, $\theta=15^\circ$ (left) and $25^\circ$ (right).}
\label{fig:compallcases}
\end{figure}

Finally, we present a comparison of the model prediction and the LES data as a function of frequency in figure \ref{fig:compallcases} for all the cases investigated in this study at two different propagation angles, $\theta=15^\circ$ and $25^\circ$ measured from the downstream end, which correspond to $x/D=18.7$ and 10.7, respectively, in previous figures showing PSD data. The region $\theta<25^\circ$ roughly determines the acoustic field dominated by the first RESPOD forcing mode and thus, the region of validity of the model. And $\theta=15^\circ$ is near the sponge zone limit used in the resolvent computations. {The frequency range in this comparison is extended to $St=[0.1,1.5]$ to show the model performance beyond the range it has been tuned for. We limit the analysis to this range since, below $St=0.1$, the hydrodynamic fluctuations reaches the acoustic field as reported by \cite{nekkanti_aiaa_2021} who used the same database for the cases M09Mc00 and M07Mc00; and at $St=1.5$, the acoustic level is already 20 dB less than the peak in all the cases.} 

For the range $St=[0.4,1]$, the model accurately predicts the acoustic field for all the cases, despite some underestimation for the cases M04Mc00, M07Mc15 and M09Mc30 at $\theta=25^\circ$. But at slightly lower propagation angles, the model starts to yield better predictions for these three cases as well, as can be seen in figures \ref{fig:respredict}, \ref{fig:coflowmodel2} and \ref{fig:coflowmodel3}. {At $St=1.5$, the model yields accurate predictions in all the cases except for the cases M07Mc00 and M08Mc00 at $\theta=15^\circ$. At frequencies below $St=0.4$, the model starts to overpredict the acoustic level which becomes evident at $St=0.1$ in all the cases. We expect the model to be valid only above a certain frequency, the forcing amplitude is scaled by $1/St^{3/2}$ in \eqref{eq:modelfinal}, which tends to infinity as $St\to 0$. The overprediction remains within 3 dB at $St=0.2$ for all the cases except M07Mc00, M07Mc15 and M09Mc30, where it reaches up to 5 dB. Another reason that may explain the poor performance at low frequencies is that the identity of the leading resolvent mode switches from Kelvin-Helmholtz to Orr mechanism at around St = 0.3 \citep{pickering_jfm_2020}, and the very different physics of Orr modes would require a different forcing model.}

\subsection{Discussion on the empirical modelling}
{We have presented a rank-1 model for acoustic sources in subsonic jets, defined within the resolvent framework. {The overall prediction involves mean flow and turbulent kinetic energy data, which can be obtained from a RANS solution.} It is known that there is a strong connection between the wavepackets found in the jet near field and the noise in the acoustic field \citep{cavalieri_jfm_2012}. It is therefore reasonable to assume that the forcing responsible for noise generation emerges from the nonlinear interaction of different wavepackets in the jet. The accuracy of the predictions implies that our model successfully incorporates the essential information from these wavepacket interactions to generate the downstream noise. The model consists of forcing structures with supersonic phase speeds determined by shifting the jet Mach number by a constant value. Such structures with supersonic phase speeds in subsonic jets are potentially a product of (\emph{i}) spatial modulation of the convected waves due to the shape of the wavepacket, yielding a supersonic tail in the wavenumber domain \citep{crighton_pas_1975,tam_jfm_2008,jordan_arm_2013}, or (\emph{ii}) jittering in the wavepackets, which manifests in the frequency domain as coherence decay, causing a shift in the energy of the wavepacket towards supersonic wavenumbers \citep{cavalieri_jsv_2011,cavalieri_jfm_2014}. Both model problems \citep{cavalieri_jfm_2014,cavalieri_amr_2019} and real jet data \citep{maia_rspa_2019,daSilva_jasa_2019} have shown that the noise generated by a wavepacket in a subsonic jet is highly sensitive to the coherence decay rate embedded in the source model. Indeed, a source model based on a wavepacket with unit coherence, although matching the near-field wavepacket obtained experimentally, generates an acoustic field that is off by up to 40 dB compared to the experimental data \citep{baqui_aiaa_2013,jordan_ctr_2014}. This implies that the supersonic structures we observe in the forcing data are more likely to be associated with the coherence decay in the jet near field, and thus the jitter mechanism. Given this perspective, the model presented here can be considered to provide an indirect representation of the coherence decay occurring due to jittering in subsonic turbulent jets.}

The present approach is based on the Mach-wave mechanism. Such a model is bound to be limited to low propagation angles, as the phase speed corresponding to $90^\circ$ should tend to infinity for the same mechanism to be responsible for side stream propagation as well. %It is already known that jet noise at $90^\circ$ is mainly defined by fine scale turbulence \citep{tam_jfm_2008}. 
However, we believe that a similar analysis based on resolvent framework can still be helpful in understanding the underlying mechanism for sidestream propagation in subsonic jets, which is left as a future task.

\section{Conclusions} \label{sec:conc}
We outlined a methodology to identify the source of subsonic jet noise at low (downstream) propagation angles. Since noise generation by turbulent flows is nonlinear, it is not possible to uniquely define the source terms.  Acoustic analogies (\citet{lighthill_prs_1952,lilley1974noise,howe_jfm_1975,doak_ap_1995,goldstein_2003}; etc.) recast the Navier-Stokes (N-S) equations as a acoustic wave equation, with all other terms considered as the source.  In this study, we instead adopt the resolvent framework, in which the linearized N-S equations serve as the operator and all nonlinear terms remaining after linearisation about the mean flow are viewed as the source terms, or \emph{forcing}, in resolvent terminology. 

Using the resolvent framework, we showed that downstream noise is generated mainly by the streamwise momentum forcing term. We then obtained a low-rank reconstruction of this forcing term using the RESPOD method \citep{towne_aiaa_2015,karban_jfm_2022}. The RESPOD method yields forcing modes that generates the SPOD modes of the measured response, which is selected to be the acoustic pressure in this study. The response modes are orthogonal to each other by construction. Searching for a similar orthogonality on the forcing side, we projected the RESPOD modes of the forcing onto streamwise harmonic waves with different phase velocities varying in the supersonic range, which yielded two critical outcomes: (\emph{i}) projection coefficients corresponding to the first RESPOD mode of the forcing peaked around the same phase velocity at all the frequencies investigated; (\emph{ii}) projection coefficients corresponding to the second RESPOD mode of the forcing showed a dip around the same phase velocity as a trace of the orthogonality in the response. Decomposing the first RESPOD mode of the forcing into supersonic and subsonic components, we demonstrated that it is the supersonic part of the forcing which generates the majority of the acoustic field in the downstream region. 

{The resolvent framework requires the mean flow data and the forcing model requires the turbulent kinetic energy, in case of non-zero flight stream. In this study, we obtained these data using the LES, which was shown to match the experimental data \citep{bres_jfm_2018,maia_arxiv_2022}. One can alternatively obtain these performing a RANS simulation, which are then expected to be less accurate. The dependency of the model results on the turbulence models used in case of a RANS simulation, and on the accuracy of the first-order statistics in general is yet to be determined in a future study.}

Given that the forcing modes of the acoustic resolvent operator, i.e., the resolvent operator that includes the measurement matrix that extracts the acoustic pressure as the response, supports a radially compact line source as the optimal forcing, we integrated the identified forcing in the radial direction, which yielded a wave packet with a dominant wavenumber corresponding to a constant phase velocity for all frequencies in the range $St=[0.4,1]$. Using this information, we introduced a model equation for the line source. We tested the line source for different flow cases with or without flight stream effects. Tuning the model by comparing the acoustic response it generated against the noise field extracted from the LES data resulted in a model in which the amplitude is scaled with $M_j^{7/2}$ and $St^{-3/2}$ and a linear phase relation is obtained changing with $St$ and $M_\infty$. The model generates a noise field with an error of less than 2 dB in the downstream region in subsonic jets over a range of frequencies.

Identifying a dominant phase speed in the acoustically efficient forcing can be of practical importance beyond yielding a source model for noise generation in subsonic jets. One can investigate the interaction mechanisms that generate forcing components at this phase speed. Given the elongated wavepacket structure observed in the source, it is reasonable to assume that these structures are associated with interaction of certain wavepackets and may potentially be traced back in the nozzle, which may help to design strategies to control the jet noise. It was already shown in \cite{maia_prf_2021} that real-time control in forced jets is possible by measuring the stochastic disturbances in the upstream region near the nozzle exit. The control practice becomes much harder in unforced jets due to loss of coherence between the actuators and the measurements. But this loss of coherence might be due to very poor signal-to-noise ratio in unforced jets. Our observation that the acoustically efficient forcing amounts to less than 0.04\% of the total forcing energy for the $M_j=0.4$ case supports this hypothesis. Extracting the structures at the dominant phase speed observed in the forcing in real-time by two-point measurements can significantly enhance the signal-to-noise ratio that is necessary for a successful control application. 

\backsection[Funding]{This work has received funding from the Clean Sky 2 Joint Undertaking under the European Union’s Horizon 2020 research and innovation programme under grant agreement No 785303. U.K. has received funding from TUBITAK 2236 Co-funded Brain Circulation Scheme 2 (Project No: 121C061). A.T. was supported in part by ONR grant N00014-22-1-2561.}

\backsection[Declaration of interests]{The authors report no conflict of interest.}

\appendix
\section{Errors in the numerical database} \label{subsec:error}

To achieve accurate resolvent-based predictions of the response, the forcing and the response data should satisfy \eqref{eq:forceinfreq}. Similarly, the RESPOD method assumes that the forcing and the response are connected to each other in the frequency domain via the resolvent operator as in \eqref{eq:resolventspodp}. The LES database contains errors from several sources that cause the above conditions to be violated. Here, we briefly discuss here these error sources and their potential effect on the results.

The LES data is generated by solving the spatially filtered N-S equations. The forcing data, as discussed in \S\ref{subsec:les}, is obtained by computing the numerical Jacobian of this nonlinear LES operator to create a consistent forcing.  The resolvent code, on the other hand, uses the linearised N-S equations without taking into account the filtering of the sub-grid scales implemented in the LES solver. This creates a compatibility issue when driving the resolvent operator with the forcing data obtained from the LES. Considering that the LES is sufficiently refined to capture the acoustic signature of the jet in an accurate way, we may assume that the differences between the LES and the N-S operators are small, at least for the scales we are interested in, and the sub-grid-scale filtering do not pose significant error in the database. 

The LES data were first generated on an unstructured grid and then interpolated onto a cylindrical grid to facilitate azimuthal decomposition. Although the cylindrical grid has a distribution similar to that of the LES grid in streamwise and radial directions \citep{bres_aiaa_2017}, interpolating the data stored in control volumes onto grid points causes interpolation errors in the data. Besides, the mesh used for the resolvent operator is not identical to the cylindrical grid used to store the LES data. The difference in the mesh requirement for the LES and the resolvent operator may vary, particularly at high frequencies, as the resolvent operator does not benefit from any sub-grid-scale filtering. Using a separate mesh for the resolvent operator requires an additional interpolation, introducing additional errors (although we expect that the errors due to this second interpolation are smaller compared to the first interpolation).

The LES data were stored with a temporal downsampling ratio of 200, yielding a sampling frequency, $St_s=12.5$, for which the Nyquist limit to avoid aliasing is given as $6.25$ \citep{nyquist_ieee_1928,shannon_ieee_1948}. A detailed analysis of the aliasing the LES database was given in \cite{karban_arxiv_2022}, where it was shown that significant aliasing was observed in the forcing data even though it is negligible in the response. As resolvent analysis is performed in the frequency domain, aliasing appears as an error source for both forcing identification based on RESPOD and the resolvent-based prediction of the response.

\begin{figure}
  \centerline{\resizebox{1\textwidth}{!}{\includegraphics{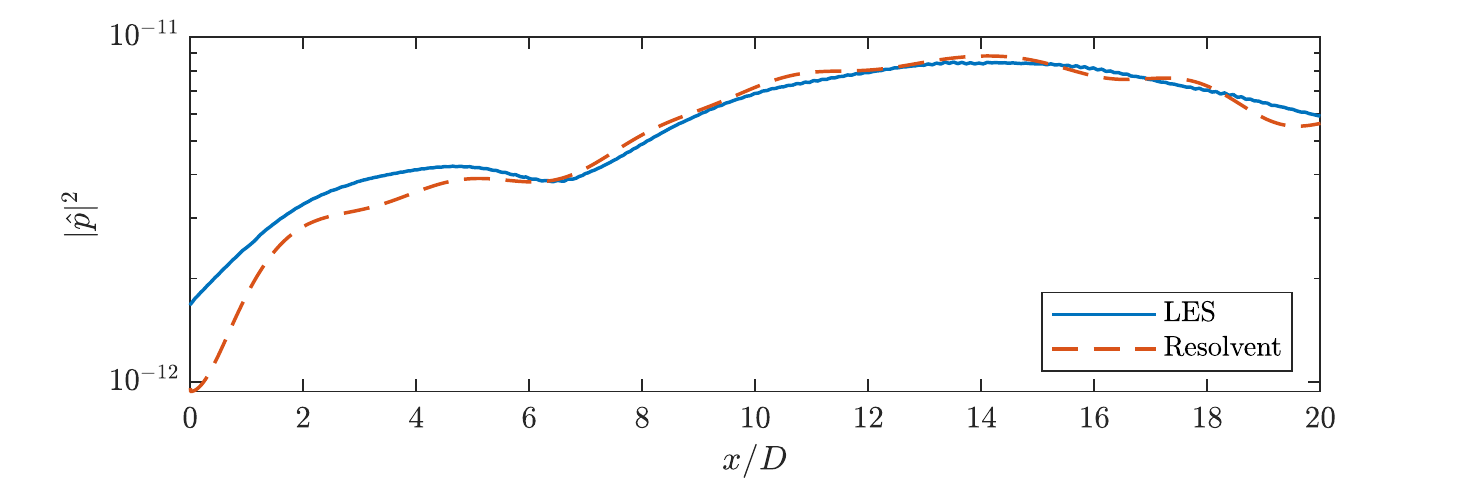}}}% Images in 100% size
  \caption{Comparison of the PSD of pressure extracted from LES and predicted via resolvent analysis at $r=5D$ for the case M04Mc00 at $St=0.6$. }
\label{fig:pcomp}
%\vspace{-15px}
\end{figure}

These errors accumulate in the LES database. It is not possible to accurately quantify contributions from each error source, but one may use the difference between the state obtained directly from the LES and its prediction obtained using the resolvent tool as a global measure of the total error included in the database. In figure \ref{fig:pcomp}, we show a comparison of the PSD of the pressure in the acoustic field, i.e., at $r=5D$, directly extracted from the LES data, and its resolvent-based prediction for the case M04Mc00 at $St=0.6$. The acoustic field is predicted with reasonable accuracy, except the most upstream part, $x<1D$, where the resolvent-based prediction suffers from a boundary condition effect.  These error levels are similar to those observed by \cite{towne_arxiv_2021} when comparing the PSD extracted from an LES database and obtained from a forced resolvent model for a supersonic jet.   These results show that, although being contaminated by errors to a certain extent, the current database can be used to investigate noise generation mechanism in jets at this Mach number.

\bibliographystyle{jfm}
\bibliography{biblio}

\end{document}